\documentclass[a4paper,12pt]{article}
\usepackage{jheppub}
\usepackage{amssymb, amsmath}

\usepackage{verbatim}
\usepackage{graphicx}
\usepackage{hyperref}
\pdfoutput=1

\newcommand{\bra}[1]{\langle #1 |}
\newcommand{\ket}[1]{| #1 \rangle}

\newcommand{\ch}{\mathop{\rm ch}\nolimits}
\newcommand{\sh}{\mathop{\rm sh}\nolimits}

\newcommand{\w}[1]{{\cal #1}}

\newcommand{\e}{\mathrm{e}}

\begin{document}
\begin{flushright}
INR-TH-2020-031\\
CERN-TH-2020-092
\end{flushright}
\vspace{-1cm}

\title{Semiclassical ${\cal S}$--matrix and black hole entropy in dilaton gravity}
\author[a,b]{Maxim Fitkevich,}
\author[a,c]{Dmitry Levkov,}
\author[d,e,a]{and Sergey Sibiryakov}
\affiliation[a]{\footnotesize Institute for Nuclear Research of the Russian Academy of Sciences, Moscow 117312, Russia}
\affiliation[b]{\footnotesize Moscow Institute of Physics and Technology, Dolgoprudny 141700, Moscow Region, Russia}
\affiliation[c]{\footnotesize Institute for Theoretical and Mathematical Physics, MSU, Moscow 119991, Russia}
\affiliation[d]{\footnotesize Institute of Physics, LPTP, Ecole
  Polytechnique Federale de Lausanne, CH-1015, Lausanne, Switzerland}
\affiliation[e]{\footnotesize Theoretical Physics Department, CERN, CH-1211 Geneva 23, Switzerland}
\emailAdd{fitkevich@phystech.edu}
\emailAdd{levkov@ms2.inr.ac.ru}
\emailAdd{sergey.sibiryakov@cern.ch}

\abstract{We use complex semiclassical method to compute
  scattering amplitudes of a point particle in dilaton gravity with a
  boundary. This model has nonzero minimal black hole mass
  $M_{cr}$. We find that at energies below 
  $M_{cr}$ the particle trivially scatters off the boundary with
  unit probability. At higher energies the
    scattering amplitude is exponentially suppressed. The corresponding
    semiclassical solution
  is interpreted as formation of an intermediate black hole decaying into the
  final-state particle. Relating the suppression of the scattering probability to
    the number of the intermediate black hole states, we find an expression
    for the black hole entropy consistent with thermodynamics. 
In addition, we fix the constant part of the entropy which is left free by the
    thermodynamic arguments. 
 We rederive this result by modifying the standard Euclidean entropy
 calculation.}
\maketitle

\section{Introduction}
\label{sec:intro}
Black hole (BH) information paradox \cite{Hawking:1976ra,Harlow:2014yka} 
has long history ever since the
discovery of BH evaporation \cite{Hawking:1974sw}. 
Recently there has
been a remarkable progress towards its resolution.
Within the framework of the AdS/CFT correspondence,
Refs.~\cite{Penington:2019npb, Almheiri:2019psf} performed 
semiclassical calculations of the entanglement entropy of an
evaporating BH and demonstrated that it follows the Page curve 
\cite{Page:1993wv,Hayden:2007cs}, consistent with unitarity.
To derive the expression for the entanglement
entropy these calculations use complex saddle points of the gravitational
path integral --- replica wormholes
\cite{Almheiri:2019qdq, Penington:2019kki, Faulkner:2013ana,
Hubeny:2007xt, Engelhardt:2014gca}.
It has been suggested that this approach 
applies also beyond the holographic setting
leading to the ``island rule'' for the entropy of the Hawking
radiation~\cite{Almheiri:2019hni,Gautason:2020tmk, Hartman:2020swn,
    Almheiri:2019qdq, Penington:2019kki}. 
It still remains to be understood, however, how quantum
correlations are encoded in the state of the emitted quanta. Only then the
information paradox will be completely
resolved {\cite{Maldacena:2001kr, Almheiri:2012rt}}. 
 
A direct approach to study unitarity of BH evaporation is to
compute the related elements of the gravitational ${\cal
    S}$--matrix~\cite{tHooft:1996rdg}. In this case one treats BH
  formation and its subsequent decay as a scattering
process~\cite{tHooft:1996rdg, Giddings:2009gj} mediated by a metastable
bound state.  
On general grounds, consideration of this complete process
  appears more adequate than its splitting into  separate stages of
collapse and evaporation. It was argued in \cite{Bezrukov:2015ufa} 
that when both initial and final states of the scattering process 
are semiclassical, the related amplitudes can be evaluated using
complex saddle points of the path integral with appropriate
  boundary conditions, cf.~\cite{Berezin:1999nn, Parikh:1999mf}. 

In this paper we further develop complex semiclassical method for
gravitational ${\cal S}$-matrix. Using this method, we compute the
  scattering amplitudes and probe the entropy of black holes in
  $(1+1)$-dimensional dilaton gravity.

We start with an outline of the method. Consider complex quantum transition
including collapse of matter in pure initial state $\Psi_i$ into a
black hole and evaporation of the latter into the state $\Psi_f$. This
process interpolates between the free flat--space states $\Psi_i$ and
$\Psi_f$ and therefore defines a gravitational ${\cal S}$--matrix
\cite{tHooft:1996rdg}. Schematically, one can write a path integral for the
transition amplitude as  
\begin{equation}\label{grav-path-int}
{\cal A}_{fi}\equiv \bra{\Psi_f}\hat{\cal S}\ket{\Psi_i}=\int{\cal
  D}\Phi\,\e^{i S'[\Phi]}\,\Psi_f^\ast[\Phi]\, \Psi_i[\Phi]\;, 
\end{equation}
where $S'[\Phi]$ is the classical action and $\Phi$ includes 
all fields of the model -- matter fields, metric, and
Faddeev--Popov ghosts.  
Precise definition of the gravitational path integral
(\ref{grav-path-int}) is a formidable task. One can assume, however,
that the initial and final states of the process are semiclassical.
In field theory this means that they contain many quanta at high
occupation numbers.
Then the integral can be evaluated in the saddle--point
approximation, giving ${\cal A}_{fi}\simeq \e^{iS'[\Phi_{cl}]}\,\Psi_f^*[\Phi_{cl}]\, \Psi_i[\Phi_{cl}]$,
where the semiclassical configuration $\Phi_{cl}$ extremizes the integrand in Eq.~\eqref{grav-path-int}
i.e. solves the classical field equations. 

Importantly, $\Phi_{cl}$ does not coincide with the classical collapsing
solution: like all configurations in the path integral
(\ref{grav-path-int}) it starts from the flat space in the past and
arrives to it in the future. Since real solutions with these properties do not
exist, $\Phi_{cl}$ is a complex saddle point describing an
exponentially suppressed process. This is to be expected: the
intermediate black hole mainly emits Hawking radiation with low
occupancies, and the probability of producing a semiclassical state
$\Psi_f$ is exponentially small. 

Generically, there may exist many complex saddle points for
  Eq.~(\ref{grav-path-int}), and one has to select 
the physical one giving the main contribution into the path
integral. To this end, we use the method suggested in 
\cite{Bezrukov:2015ufa} (see
\cite{Bezrukov:2003tg, Levkov:2007yn} for quantum mechanical
applications). 
The main
idea is to enforce the scattering boundary conditions in the path
integral (\ref{grav-path-int}) 
with a special variant of a constrained instanton method. 
After that
the physical complex solutions are obtained by smooth deformation
of the real solutions that describe classical low-energy 
scattering without black hole production. 

Our method reduces construction of the semiclassical gravitational
${\cal S}$--matrix to solution of the 
classical field equations in the complex
domain. Though this is in principle managable,  
applications to four--dimensional field theories with
dynamical gravity are challenging. 
So far this method has been applied only in
spherically reduced models
with simplified matter content \cite{Bezrukov:2015ufa}.

\begin{figure}[ht]
\begin{center}
\center{\includegraphics{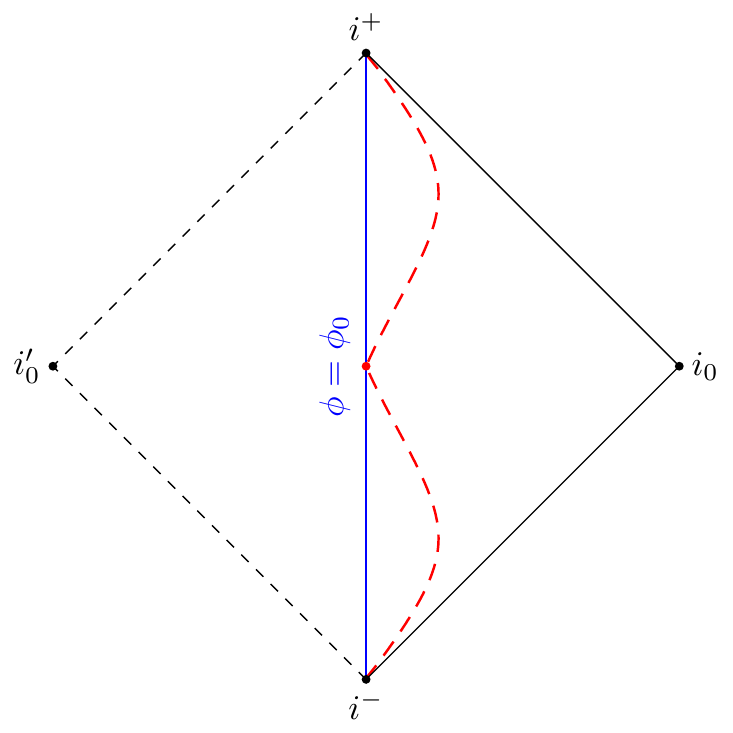}}
\caption{Penrose diagram for the vacuum solution in the CGHS model. The
  boundary $\phi=\phi_0$ cuts off the strongly coupled region to the
  left making the model semiclassically tractable. Dashed line shows
  the trajectory of a particle reflecting off the
  boundary. } \label{fig:1} 
\end{center}
\end{figure}

Below we consider another simplified model based on 
the two--dimensional
Callan--Giddings--Harvey--Strominger (CGHS)
\cite{Callan:1992rs, Strominger:1994tn} dilaton gravity.
The model describes interaction of a non--dynamical
metric $g_{\mu\nu}(x)$ and dilaton $\phi(x)$ with matter.
The action of this model is qualitatively similar to that of
  spherically--reduced multidimensional gravity, where $g_{\mu\nu}$
  includes the time and radial metric components and
  $\mathrm{e}^{-2\phi}$ is related to the areas of the extra
  spheres~\cite{Strominger:1994tn}.
The vacuum solution in this model has flat $g_{\mu\nu}$ and linear dilaton
field $\phi$ changing from $-\infty$ to $+\infty$. For
positive values of $\phi$, gravity becomes strongly coupled
precluding the semiclassical analysis.   
To make the model tractable, we cut off the strongly coupled region by
introducing 
a reflective boundary along the
line of constant dilaton $\phi(x)=\phi_0$, where $\phi_0$ is
negative and large
\cite{Russo:1992ax, Chung:1993rf, Strominger:1994xi, Das:1994yc,
  Fitkevich:2017izc}.
All fields in the path integral are then restricted 
to the submanifold $\phi<\phi_0$ (the rightmost region in
Fig.~\ref{fig:1}). 
This model was shown to be equivalent to the flat limit of the
Jackiw--Teitelboim gravity
\cite{Teitelboim:1983ux, Jackiw:1984je}
with a boundary both at the classical \cite{Cangemi:1992bj} and
quantum level \cite{Fitkevich:2020okl}.

We also make the second radical
simplification. 
Instead of a full-fledged field theory, we represent the matter sector
with 
a point particle of mass $m$ moving along the trajectory  
$x^\mu = x^\mu_*(\tau)$.
One can interpret it as a toy model for the narrow wavepacket in field
theory. 
We find complex semiclassical solutions
${\Phi_{cl}=\{g_{\mu\nu}(x),\phi(x),x^\mu_*(\tau)\}}$ and compute the
transition amplitudes of the particle. At low energies $M$ the
particle trivially scatters off the boundary with unit probability, see the dashed line in Fig.~\ref{fig:1}. However, once the energy exceeds
a certain critical value $M_{cr}$ the semiclassical solutions become
complex. Initial and final parts of these solutions describe formation of an
intermediate BH with mass $M$ from the particle and,
after complex evolution, a 
particle in the final state. The transition probability equals
\begin{equation}\label{trans-prob}
{\cal P}_{fi}=|{\cal A}_{fi}|^2 \simeq
\e^{-2\pi(M-M_{cr})/\lambda }\;, \qquad\qquad M>M_{cr}\;,
\end{equation}
independently of the particle mass $m$. Here $\lambda$ is the CGHS
energy scale. Notably, $M_{cr}$ coincides with the minimal mass of black holes in
the model.
It is worth stressing that our semiclassical
method provides the phase of the amplitude, in addition to its
absolute value. 

One can interpret the probability \eqref{trans-prob} as
follows~\cite{Parikh:1999mf}. The intermediate BH has entropy
$\Sigma_{BH}(M)$ and 
an exponentially
large number of states $\exp(\Sigma_{BH})$.
Then it is expected to decay into the
single-particle final state with probability ${{\cal
  P}\propto\exp(-\Sigma_{BH})}$. Comparing to
Eq.~\eqref{trans-prob}, we find the entropy of the CGHS black
  holes, 
\begin{equation}\label{corr-entr}
\Sigma_{BH}=2\pi (M-M_{cr})/\lambda\;.
\end{equation}
This expression is consistent with the results for BH entropy
 in similar 
models~\cite{Fiola:1994ir,Myers:1994sg,Hayward:1994dw,Solodukhin:1995te}. 

Our result, however, raises a puzzle.
A naive extrapolation to our model of the 
Gibbons-Hawking Euclidean calculation~\cite{Gibbons:1976ue} of
the BH
entropy gives,
\begin{equation}
\label{Snaive}
\Sigma_{BH}^{naive}=2\pi M/\lambda\;,
\end{equation}
independently of the boundary parameter
$\phi_0$. The expression~\eqref{Snaive} would imply 
that the entropy of the critical black hole with mass $M_{cr}$ is
non-zero. If this were the case, one would see an unphysical jump of the
scattering probability ${\cal P}_{fi}$ at $M = M_{cr}$. Our
result in Eq.~(\ref{trans-prob}), quite consistently, has no 
jump.

Note that the constant term in BH entropy is not fixed by the laws of BH
thermodynamics. In previous Euclidean calculations of BH entropy in
dilaton gravity, this constant was added somewhat ad hoc. We show that
Eq.~(\ref{corr-entr}) 
can be recovered naturally by a suitable modification of the
Euclidean procedure once the presence of the boundary
at $\phi=\phi_0$ is taken into account. 

It is worth stressing that the arguments leading to Eq.~(\ref{corr-entr}) 
do not apply to multidimensional gravity, where critical BHs are known 
to have
nonzero entropy \cite{Gibbons:1976ue, Strominger:1996sh}. 
The  masses of the latter are minimal only  
among the black holes with given charges and/or angular momenta,
whereas the absolute minimum is reached by the neutral BH with
the Planckian mass. In this case collision
  of charged particles may lead to formation of a neutral BH, with
  charge and angular momentum carried away by
  bremsstrahlung. Then the corresponding scattering probability is a
  continuous function of energy~\cite{Bezrukov:2015ufa}. 

The present paper is organized as follows. 
In
Sec.~\ref{sec:setup} we introduce our setup. The scattering amplitude
is calculated in 
Sec.~\ref{sec:s-matrix}.  
In Sec.~\ref{sec:suppression} we discuss the entropic interpretation
of the scattering probability and the 
Euclidean calculation of BH entropy. 
Section~\ref{sec:conclusion} is devoted to 
conclusions. Several Appendices contain details of the calculations.

\section{The setup}
\label{sec:setup}
\subsection{Dilaton gravity}
\label{sec:action}

We consider non--perturbative scattering in two--dimensional dilaton
gravity with a boundary \cite{Fitkevich:2017izc},
see also
\cite{Callan:1992rs, Chung:1993rf, Das:1994yc, Strominger:1994xi,
Russo:1992ax, Fitkevich:2020okl}. The
gravitational action\footnote{We use the metric signature $(-,+)$ and
  Greek indices $\mu,\nu,\ldots=0,1$.}
\begin{gather}\label{eq:grav-action}
S_{{gr}}=\int\limits_{\phi<\phi_0}d^2x\,\sqrt{-g}\, 
\e^{-2\phi}\left[R+4(\nabla\phi)^2+4\lambda^2\right]
+2\int\limits_{\phi=\phi_0}d\tau_0\,\e^{-2\phi}\left(K+2\lambda\right)
\end{gather}
describes the CGHS model \cite{Callan:1992rs} with non--dynamical metric
$g_{\mu\nu}(x)$ and dilaton $\phi(x)$. Besides, it includes the
timelike boundary at $\phi=\phi_0$ which cuts off the region of strong
coupling. Importantly, a regulating boundary should be present in all
configurations in the path integral \eqref{grav-path-int}, otherwise
the CGHS fields would become singular at the quantum level
\cite{deAlwis:1992emy, Fitkevich:2020okl}. In
Eq.~\eqref{eq:grav-action} we included the Gibbons--Hawking term
\cite{Gibbons:1976ue} at $\phi=\phi_0$ with proper time $\tau_0$,
extrinsic curvature $K=\nabla_\mu n_0^\mu$ and outer
normal\footnote{The direction of the normal is fixed by the condition
  $n_0^\mu\nabla_\mu \phi>0$.}
$n_0^\mu$. Parameter $\lambda$ sets the energy scale of the
model. 

The semiclassical expansion is controlled by the combination 
$\e^{2\phi_0}$. Indeed, a shift $\phi\mapsto\phi+\phi_0$ brings this
parameter in front of the classical action, at the place of the Planck
constant in the path integral. In what follows we consider the case 
\begin{equation}\label{eq:5c}
\e^{2\phi_0}\ll 1\;,
\end{equation}
and work to the leading order in this parameter.

\begin{figure}[ht]
\begin{center}
\center{\includegraphics{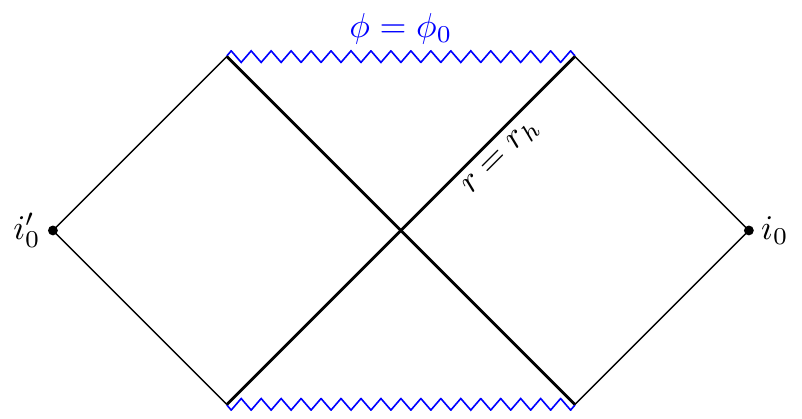}}
\caption{Black hole in the CGHS model with a boundary. The field equations
   break down at the line $\phi=\phi_0$, indicating a singularity.} \label{fig:2}
\end{center}
\end{figure}

Without matter, the general solution 
in the bulk is,
\begin{equation}\label{sol:black}
ds^2=-f(r)\, dt^2+\frac{dr^2}{f(r)}\;, \qquad \phi=-\lambda r\;, \qquad f(r)=1-\frac{M}{2\lambda}\, \e^{-2\lambda r}\;,
\end{equation}
where $M$ is the Arnowitt--Deser--Misner (ADM) mass.
This constitutes the two-dimensional analog on the Birkhoff
theorem~\cite{LouisMartinez:1993cc}, which we derive in  
Appendix~\ref{sec:equations} for completeness. 
For $M=0$ the spacetime is flat, while for $M>0$ it describes a black hole.
In Eqs.~\eqref{sol:black} we use Schwarzschild coordinates with the ``radius''
$r=-\phi/\lambda$ and the orthogonal time $t$. The light--like line $r=r_h$,
\begin{equation}\label{eq:rh}
r_h=\frac{1}{2\lambda}\log\left(\frac{M}{2\lambda}\right)\;,
\end{equation}
with $f(r_h)=0$ is a black hole horizon. Penrose diagrams of the
solutions with $M=0$ and $M>0$ are shown in
Figs.~\ref{fig:1} and~\ref{fig:2}, respectively.

It is not enough, however, to solve the bulk field equations: one
should also add the boundary. 
This amounts to cutting off the spacetime at $\phi=\phi_0$ 
and imposing the boundary condition
\begin{equation}\label{dil-bc}
n_0^\mu \nabla_\mu\phi=\lambda\;\qquad \qquad\text{at}\;\;\;\phi=\phi_0\;,
\end{equation}
which follows from variation of the action \eqref{eq:grav-action} with
respect to the boundary metric, see Appendix~\ref{sec:refl-laws}.
The spacetime \eqref{sol:black} 
satisfies Eq.~\eqref{dil-bc} only for $M=0$ when it is flat.
The breakdown of the equations of motion at the line $\phi=\phi_0$ for 
$M\neq0$ 
implies that it should be interpreted as a singularity.
This line is
spacelike and hidden under the black hole horizon if $r_h>-\phi_0/\lambda$
or $M>M_{cr}$, where 
\begin{equation}
M_{cr}=2\lambda \e^{-2\phi_0}
\end{equation}
is the critical
mass, see Fig.~\ref{fig:2}. 
At $M<M_{cr}$, $M\neq 0$ the solution \eqref{sol:black} is a spacetime with
timelike naked singularity. The latter 
does not form in the collapse of a regular
matter~\cite{Fitkevich:2017izc}. 

\subsection{Classical scattering}
\label{sec:cl-scat}
Now we want to consider scattering of a point particle with
mass $m$ and action 
\begin{equation}\label{eq:partic-action}
S_m=-m\int d\tau
\end{equation}
off the boundary. 
Here the parameter $\tau$ is a
proper time of the particle. 
One can find the particle trajectories using the well--known techniques
developed for thin shells in multidimensional gravity
\cite{Berezin:1987bc}. We describe the particle trajectory with
  radius $r = -\phi/\lambda$ as a function of the
  proper time $r=r_*(\tau)$. The two--dimensional Birkhoff theorem
guarantees 
that the empty spacetime
regions to the left and to the right of the particle are either 
Schwarzschild or Minkowski. 

Then, if the particle
starts evolution in Minkowski spacetime, the solution in the ``inner'' region
$r<r_*$
remains flat, 
\begin{equation}\label{sol:flat}
ds^2=-dT^2+dr^2\;, \qquad\qquad \phi=-\lambda r\;.
\end{equation}
Note that we introduced the notation $T$ for the time
coordinate in the inner region to emphasize its
difference from the time $t$ of the distant observer. 
Similarly, the ``outer''
region $r>r_*$ is described by the Schwarzschild metric
\eqref{sol:black} with conserved gravitational mass $M$. 

Since the particle energy--momentum tensor is concentrated at the
worldline, the derivatives of the metric and dilaton 
change discontinuously across
it. In Appendix~\ref{sec:junction} we derive the Israel junction
conditions for the jumps of the extrinsic curvature and normal
derivative of the dilaton, 
\begin{equation}\label{israel}
[n^\mu\nabla_\mu\phi]=\frac{m}{4}\,\e^{2\phi(r_*)}\;,\quad \qquad
[K]=2\,[n^\mu\nabla_\mu\phi]\;. 
\end{equation}
Here the square brackets represent difference of the values at $r_*+0$ and
$r_*-0$; the worldline normal $n^\mu$ points towards large $r$. 
Substituting the inner and outer spacetimes \eqref{sol:flat},
\eqref{sol:black} into Eq.~\eqref{israel} one finds equation of
motion for the particle, 
\begin{equation}\label{eq:effective_particle}
\dot{r}_*^2+V_{\mathrm{eff}}(r_*)=0\;, \qquad\qquad
V_{\mathrm{eff}}(r)=1-\left(\frac{M}{m}+\frac{m}{8\lambda}\,\e^{-2\lambda
    r}\right)^2\;, 
\end{equation}
where dot is a derivative with respect to the proper time
$\tau$.
Recall that $M > m$ is the particle total energy measured at
  infinity, cf.\ Eq.~\eqref{eq:particle-eq-app}.
This equation has an intuitive form
of non--relativistic ``energy conservation law'' with effective
potential $V_{\mathrm{eff}}$. The latter is negative everywhere, it
monotonically
increases from a finite value at the boundary $r=r_0$,
\begin{equation}
r_0=-\phi_0/\lambda\;,
\end{equation}
to
$1-M^2/m^2<0$ as $r\to+\infty$. The details of the derivation are
given in Appendix~\ref{sec:junction}. 

Now it is clear that the left--moving
particle with energy $M<M_{cr}$ always reaches the boundary
$r=r_0$ at some moment of time $\tau=\tau_\times$.
Then it reflects back. 
In Appendix~\ref{sec:refl-laws} we demonstrate that 
reflection of the
particle from the boundary simply flips the sign
of its radial velocity,
$\dot{r}_*(\tau_\times+0)=-\dot{r}_*(\tau_\times-0)$. At
late times the particle goes to $r\to+\infty$.  

The classical story changes completely if the particle energy $M$
exceeds $M_{cr}$. In this case it first crosses the horizon $r_h>r_0$
of the outer metric \eqref{sol:black} and thus forms a black hole. 
Whence the particle can be retrieved only quantum
mechanically with exponentially
small probability.  

\section{Semiclassical scattering amplitude}
\label{sec:s-matrix}
\subsection{Semiclassical method}
\begin{figure}[ht]
\begin{center}
\center{\includegraphics{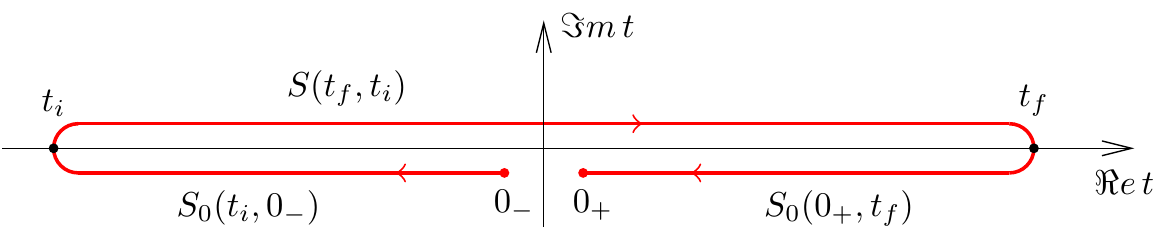}}
\caption{ Time contour in  the path integral for the scattering
  amplitude.} 
\label{fig:new-contour} 
\end{center}
\end{figure}

Quantum ${\cal S}$--matrix is an operator connecting initial and final Fock
states of the process. It is formally defined as 
\begin{equation}\label{s-matrix}
\hat{\cal S}=\hat{U}_0(0,t_f)\,\hat{U}(t_f,t_i)\,\hat{U}_0(t_i,0)\;,
\end{equation}
where $\hat{U}$ and $\hat{U}_0$ are the interacting and free evolution
operators, and the limits $t_i\to-\infty$, $t_f\to+\infty$ are
assumed. In the path integral representation Eq.~\eqref{s-matrix}
reads, 
\begin{equation}\label{amplitude}
\w{A}_{fi}\equiv\bra{\Psi_f}\hat{\cal S}\ket{\Psi_i}=
\int {\cal D}\Phi\,\e^{iS_0(0_+,t_f)+iS(t_f,t_i)+iS_0(t_i,0_-)}\,\Psi_f^\ast[\Phi]\,\Psi_i[\Phi]\;,
\end{equation}
where $\Phi$ denotes all fields of the model on the time
contour in Fig.~\ref{fig:new-contour}, while $S$ and $S_0$ are the
interacting and free classical actions\footnote{We shortly denoted
  $S'[\Phi]\equiv S_0(0_+,t_f)+S(t_f,t_i)+S_0(t_i,0_-)$ in
  Eq.~\eqref{grav-path-int}.} on the respective parts of the
contour. Note that the fields at the endpoints of the contour $t=0_-$
and $t=0_+$ do not coincide. 
We also introduced the wave functions $\Psi_i$, $\Psi_f$ of the free
initial and final states. 

We generalize Eq.~\eqref{amplitude} to gravity in a straightforward
way. In this case $\Phi$ includes the particle trajectory $r_*(\tau)$,
metric $g_{\mu\nu}$, dilaton $\phi$ and Faddeev--Popov ghosts. The
interacting action
\begin{equation}\label{eq:action-sum}
S(t_f,t_i)=S_{gr}+S_m+S_{GH}\;,
\end{equation}
involves gravitational and matter contributions, as well as the
standard Gibbons--Hawking term $S_{GH}$ at infinity; see its
definition in Appendix~\ref{sec:grav-act}. 

Importantly, we assume that configurations
$\Phi=\{g_{\mu\nu}(x),\phi(x),r_*(\tau)\}$ in the path integral
\eqref{amplitude} have trivial topology 
expected from the scattering processes. First, they should
contain the boundary $\phi=\phi_0$. Second, they should
start from flat spacetime at $t=t_i$ and arrive to it in the
future. This gives a preferred choice of the asymptotic time $t$
changing from $t_i$ to $t_f$ by the clock of the distant observer. The
free actions $S_0=S_m$ describe a particle in flat spacetime
and $\Psi_{i,f}\propto \e^{\mp ipr}$ are the momentum eigenstates of
this particle with $p=\sqrt{M^2-m^2}$. 

In the semiclassical limit $\e^{2\phi_0}\ll 1$ the classical action $S$
becomes large and the integral \eqref{amplitude} can be evaluated in
the saddle--point approximation,
\begin{equation}\label{eq:3.3}
\w{A}_{fi}\simeq \e^{iS_{tot}[\Phi_{cl}]}\;,
\end{equation}
where
\begin{equation}\label{eq:total}
S_{tot}[\Phi]=S_0(0_+,t_f)+S(t_f,t_i)+S_0(t_i,0_-)-i\ln\,\Psi_f^\ast-i\ln\,\Psi_i
\end{equation}
is the total action and $\Phi_{cl}$ is a complex classical solution
extremizing $S_{tot}$. The Faddeev--Popov ghosts can be neglected at
this point as they don't contribute to the leading exponential term.

\subsection{From low to high energies}
It is straightforward to compute the amplitude at $M<M_{cr}$ substituting the
real classical solution into
Eq.~\eqref{eq:3.3}, see Appendix~\ref{sec:grav-act}. 
In the overcritical case, however, the task of finding the relevant
saddle--point configuration becomes non-trivial.
The ordinary
collapsing solutions are of no use here, since they describe formation
of black holes and therefore violate the requirement of flat spacetime
in the asymptotic future. 

To enforce this requirement, we introduce 
a positive--definite and
diffeomorphism--invariant functional 
$\w{T}_{\mathrm{int}}[\Phi]$ 
estimating the duration of the
scattering process from the viewpoint of a distant observer. Namely,
$\w{T}_{\mathrm{int}}[\Phi]$ should be finite on any scattering
configuration $\Phi$ interpolating between flat spacetimes at
$t\to\pm\infty$, and infinite otherwise. 
Then we constrain the path integral (\ref{amplitude}) to run only over
configurations with finite values of
$\w{T}_{\mathrm{int}}[\Phi]$. Technically, this is implemented by 
inserting the unity
\begin{equation}\label{faddev}
1=\int\limits_0^{+\infty} d\w{T}_0\;\delta(\w{T}_{\mathrm{int}}[\Phi]-\w{T}_0)=\int\limits_0^{+\infty}d\w{T}_0\int\limits_{-i\infty}^{+i\infty}\frac{d\varepsilon}{2\pi i}\;\e^{-\varepsilon(\w{T}_{\mathrm{int}}-\w{T}_0)}
\end{equation}
into the integrand of Eq.~\eqref{amplitude}
and interchanging the order of integration over ${\cal D}\Phi$ 
and~$d\w{T}_0d\varepsilon$. 
 
To have a specific example, consider the choice
\begin{equation}\label{eq:int-time}
\w{T}_{\mathrm{int}}[\Phi]=\int d^2x\,\sqrt{-g}\,L(\phi)\left[\lambda^2-(\nabla\phi)^2\right]^2\;, \qquad L(\phi)=\e^{-4\phi}\delta(\phi-\phi_\varepsilon)/\lambda^2\;,
\end{equation}
where the integration is 
concentrated on the line $\phi=\phi_\varepsilon$ which is
far away from the boundary,
$|\phi_\varepsilon|\gg|\phi_0|$. Clearly, $\w{T}_{\mathrm{int}}$
in Eq.~\eqref{eq:int-time} is positive--definite for real
  $g_{\mu\nu}$ and $\phi$. Besides, in the
asymptotically Schwarzschild spacetime with mass $M$ one finds
$\w{T}_{\mathrm{int}}=\int dt\,M^2/4\lambda$. Thus, this functional
estimates the asymptotic time spent by the ADM mass $M$ in the
``interaction region'' to the left of $\phi=\phi_\varepsilon$. We
stress that our method is not specific to the choice
(\ref{eq:int-time}) and
can exploit any appropriate positive--definite
$\w{T}_{\mathrm{int}}$. 

Inserting the unity \eqref{faddev} into Eq.~\eqref{amplitude}, one
finds the path integral 
with the ``regularized'' interacting action
\begin{equation}\label{reg-action}
S_\varepsilon[\Phi]=S[\Phi]+i\varepsilon\,\w{T}_{\mathrm{int}}[\Phi]
-i\varepsilon\,\w{T}_0
\end{equation}
and the additional integrations over $\varepsilon$ and $\w{T}_0$. The
$\delta$-function (\ref{faddev}) ensures 
that the configurations $\Phi$ leave the ``interaction
region'' in a finite ``time'' $\w{T}_0$. Besides, 
we can use the positive definiteness of 
$\w{T}_{\mathrm{int}}$, to improve convergence of the path
integral. 
To this end, we deform the contour of $\varepsilon$-integration into
the region $\Re
e\,\varepsilon\geq 0$. 

At fixed $\w{T}_0$ and $\varepsilon$ the semiclassical solutions extremize
$S_\varepsilon[\Phi]$. The additional saddle--point integrals with
respect to $\varepsilon$ and $\w{T}_0$ give $\varepsilon=0$. We therefore
perform calculations at $\varepsilon>0$ and send $\varepsilon\to+0$ in
the end, restoring the original saddle--point equations.
The ``regularized''
semiclassical solutions at $\varepsilon>0$ have three important
properties \cite{Bezrukov:2015ufa, Bezrukov:2003tg, Levkov:2007yn}. 
First, they leave the
``interaction region'' $\phi>\phi_\varepsilon$ in finite time. Second,
the corresponding fields are generically complex-valued.
Third, they can be obtained by
smooth deformation of the classical reflecting solutions. 

\begin{figure}[ht]
  \hspace{3.5cm}(a) \hspace{5.5cm} (b)

  \centerline{\includegraphics{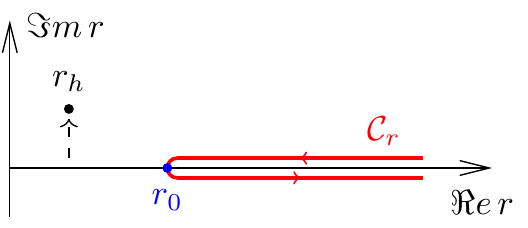}
    \hspace{1cm}
    \includegraphics{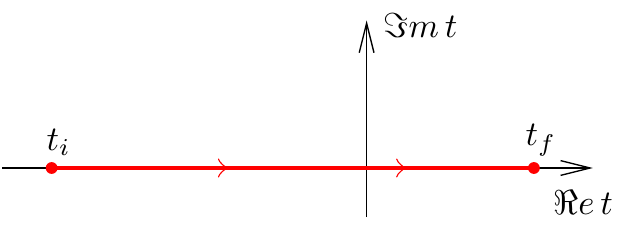}}

  \vspace{3mm}
  \hspace{3.5cm}(c) \hspace{5.5cm} (d)
  \vspace{-3mm}
  
  \centerline{\includegraphics{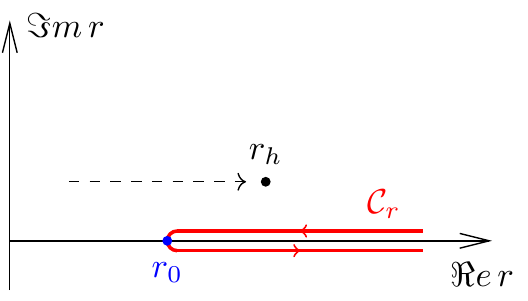}
    \hspace{1cm}
    \includegraphics{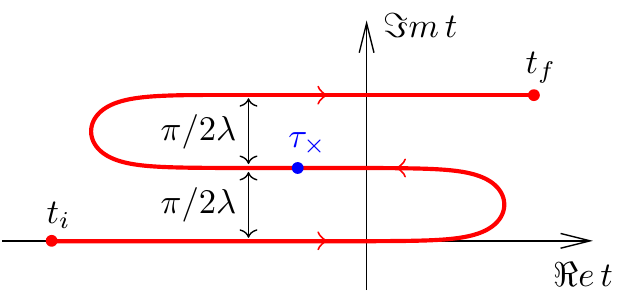}}

\caption{ Trajectory of the particle 
in complex planes of the radial and temporal 
Schwarzschild coordinates for the regularized
  solutions at $M<M_{cr}$ (top) and $M>M_{cr}$ (bottom).
The radial coordinate $r_*(\tau)$ varies along the almost real contour
${\cal C}_r$ as the particle's proper time $\tau$ changes from $-\infty$ to
$+\infty$. The particle bypasses the event horizon which is shifted upwards in the
complex plane. 
} \label{fig:contours}
\end{figure}

To demonstrate these properties, we consider the ``shell--like'' term
\eqref{eq:int-time} concentrated at $\phi=\phi_\varepsilon$. Junction
at this shell changes the metric to the left of the shell, at
$r < -\phi_\varepsilon/\lambda$. By Birkhoff theorem, the form of this metric is
still Schwarzschild, Eq.~\eqref{sol:black}, but with the complex mass 
\begin{equation}\label{complex-mass}
M\,\mapsto\,M_\varepsilon=M+i\,\varepsilon'\;.
\end{equation}
In Appendix~\ref{sec:regul-method} we show that $\varepsilon'$ is
positive and proportional to $\varepsilon$. 
After this replacement the regularized saddle--point configurations
change continuously with energy. 
At $M<M_{cr}$
they are close to the real classical solutions: the particle
trajectory $r_*(\tau)$ reaches the boundary at $r=r_0$ and reflects from
it, see Fig.~\ref{fig:contours}a. The outer time $t$ changes
almost along the 
real axis (Fig.~\ref{fig:contours}b).
Importantly, the horizon of the outer
metric now acquires a positive imaginary part, $\Im m\, r_h>0$, see
Eq.~(\ref{eq:rh}). Thus, even at $M>M_{cr}$ the particle continues to
evolve along the contour ${\cal C}_r$ in Fig.~\ref{fig:contours}c. 
It bypasses the horizon in complex
$r$--plane, both on the way in and on the way out. 
But now the outer Schwarzschild time of the particle is
  essentially complex. Equations~\eqref{sol:black} and
  \eqref{eq:effective_particle} imply,
\begin{equation}\label{time-cont}
t(r_*)=\int_{r_i}^{r_*} dr\,\frac{\sqrt{f(r)-V_{\mathrm{eff}}(r)}}{f(r)\,\dot{r}_\ast(r)}\;, \qquad \dot{r}_\ast(r)=\mp\sqrt{-V_{\mathrm{eff}}(r)}\;,
\end{equation}
where the integral runs along the contour $\w{C}_r$ 
in Fig.~\ref{fig:contours}c and the minus (plus) sign of
$\dot{r}_\ast$ correspond to motion prior to (after) reflection at
$r_0$. 
The integrand in Eq.~\eqref{time-cont} has a pole at the
horizon giving an imaginary time change 
\begin{equation}\label{eq:imaginary_time}
\Im m\,(t_f-t_i)=2\pi\,\underset{~r=r_h}{\mathrm{Res}}\,f^{-1}(r)=\frac{\pi}{\lambda}\;.
\end{equation}
Notably, the time contour in Eq.~\eqref{time-cont} is smooth at finite
$\varepsilon>0$, see Fig.~\ref{fig:contours}d. 
Since the regularized solutions are now connected to the classical
ones, we assume that they also represent the physical saddle points of
the path integral \eqref{amplitude}.\footnote{This assumption 
has been confirmed in quantum--mechanical systems by direct comparison
with the solutions of the Schr\"odinger equation 
\cite{Bezrukov:2003tg, Levkov:2007yn, Levkov:2007ce, Levkov:2008csa}.} 
Once the amplitude \eqref{eq:3.3}
is computed, we send $\varepsilon\to+0$. 

\subsection{The result}
\label{sec:supercrit-scatt}
\begin{figure}[t]
  
\centerline{\includegraphics{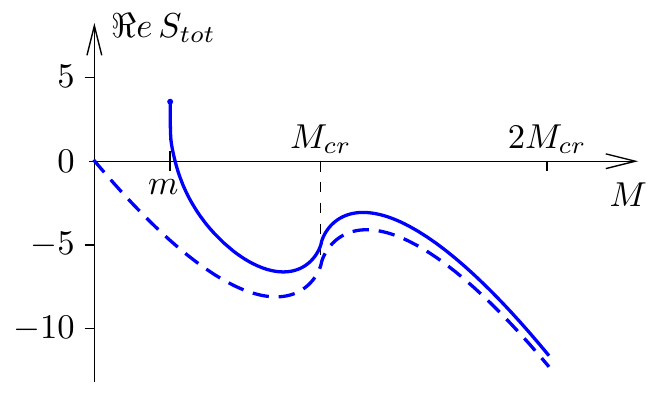}\hspace{1cm}
  \includegraphics{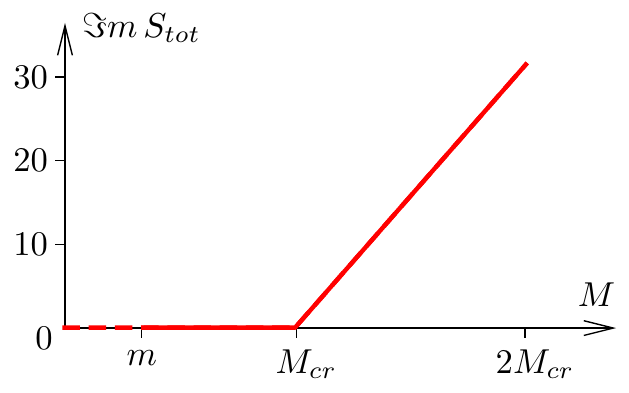}}

\hspace{3.5cm}(a) \hspace{7.5cm}(b)

\caption{ Real (a) and imaginary (b) parts of the total
action \eqref{fully-total} for $m=M_{\mathrm{cr}}/3$ (solid) and $m=0$
(dashed) as functions of the particle energy $M$.
The critical black hole mass is $M_{\mathrm{cr}}=10\lambda$. 
The interval $M<m$ is kinematically forbidden.
 } \label{fig:action}
\end{figure}

By construction, the regularized saddle--point configurations have
trivial topology, just like the reflective classical solutions at low
energies.
Their action $S_{tot}$ is computed in a straightforward way,
given Eq.~(\ref{eq:effective_particle}) for the particle
trajectory $r_\ast(\tau)$  and the inner and outer metrics 
 \eqref{sol:black}, \eqref{sol:flat}. We perform this computation in 
 Appendix~\ref{sec:grav-act}. Here is the result,
\begin{equation}
\label{fully-total}
\begin{split}
S_{tot}=&-\frac{M-M_{cr}}{\lambda}\log\left(1-\frac{M+i\varepsilon'}{M_{cr}}
\right) + \frac{p}{\lambda}(1+2\phi_0) \\
&-\frac{p}{\lambda}\log\left(\frac12+\frac{m^2
    M}{8M_{cr}p^2}+\frac{p_\times}{2p}\right) +
\frac{2M_{cr}}{\lambda}\log\left(\frac{4M_{cr}(p_\times+M)+m^2}{4M_{cr}(p_\times+M)-m^2}\right)\\ 
  & 
+\frac{M}{\lambda}\log\bigg[\frac{4M^3-3m^2M
+(4M^2-m^2)p_\times}{(p+M)^3} + 
  \frac{m^2(4M^2+m^2)}{4M_{cr}(p+M)^3}\bigg],
\end{split}
\end{equation}
where
\begin{equation}
\label{pcross}
p_\times=\sqrt{(M+m^2/4M_{cr})^2-m^2}\;
\end{equation}
is the radial momentum of the particle
immediately after the collision with the
boundary. 
This result is finite and valid at all energies.
It provides the absolute value and the phase of the amplitude
\eqref{eq:3.3}.
In the massless case $m=0$ the expression \eqref{fully-total} simplifies, 
\begin{equation}\label{eq:massless}
S_{{tot}}=-\frac{M-M_{\mathrm{cr}}}{\lambda}\,\log\,\left(1-\frac{M+i\varepsilon'}{M_{\mathrm{cr}}}\right)+\frac{M}{\lambda}\left(1-\log\,\frac{M_{\mathrm{cr}}}{2\lambda}\right)\;.
\end{equation}
The infinitesimal mass shift $i\varepsilon'$ in
Eqs.~(\ref{fully-total}), (\ref{eq:massless}) 
fixes the branch of the first logarithm at $M>M_{cr}$ leading 
to the imaginary part,
\begin{equation}\label{imag-action}
\Im m\; S_{tot}=\frac{\pi}{\lambda}(M-M_{cr})\;\theta(M-M_{cr})\;,
\end{equation}
which is independent of the particle mass $m$. This gives the
probability of overcritical scattering
$\w{P}_{fi}=|\w{A}_{fi}|^2$ in Eq.~\eqref{trans-prob}. 
The real and imaginary parts of the expression (\ref{fully-total})
are shown in Fig.~\ref{fig:action}.

\begin{figure}[ht]
\begin{center}
\center{\includegraphics{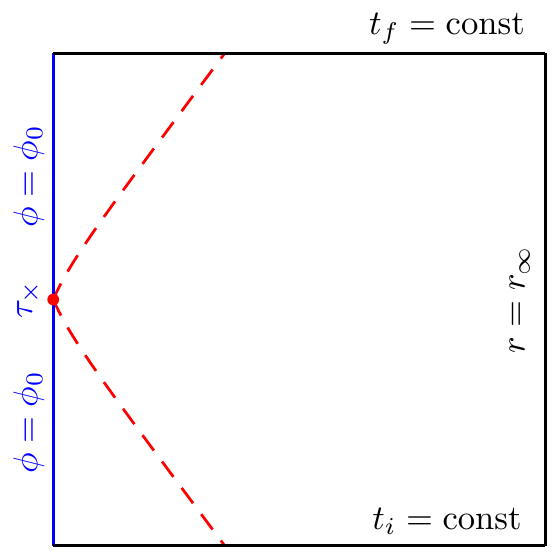}}
\caption{Schematic representation of the regularized scattering
solution. Red dashed line shows the particle trajectory.} \label{fig:4}
\end{center}
\end{figure}

Let us outline where the imaginary part of $S_{tot}$ comes from. We
consider\footnote{The analytic integral $S_{tot}$ is independent
  of the choice of the complex contour. 
However, separate contributions to it depend
on this choice.} the space--time contour in Figs.~\ref{fig:contours}c,d
with almost real $r_\ast(\tau)$. Then the particle action
\eqref{eq:partic-action} is almost real as well by
Eq.~\eqref{eq:effective_particle}. In Appendix~\ref{sec:grav-act} we
show that the bulk CGHS Lagrangian in Eq.~\eqref{eq:grav-action} is a
total derivative. This means that the interacting action 
$S$ is a sum of integrals along the
boundary $\phi=\phi_0$, initial and final Cauchy surfaces
$t=t_i,t_f$, and the timelike surface at the
spatial infinity $r=r_\infty$, see Fig.~\ref{fig:4}. 
We find that the latter term vanishes.
The integrals at $t=t_i,t_f$ combined with
the free action 
$S_0$ and the wavefunctions $\Psi_i$, $\Psi_f$ give 
complex contribution into $S_{tot}$. Its imaginary part comes from 
the residue of the
Schwarzschild time at the horizon, 
\begin{equation}\label{imag-time}
\Im m\,S_{tot}^{(1)}=M\Im m(t_f-t_i)=\frac{\pi}{\lambda} M\,\theta(M-M_{cr})\;.
\end{equation}
This is similar to the results of the previous studies 
\cite{Parikh:1999mf, Bezrukov:2015ufa}. 
Remarkably, the contribution of the boundary is also complex. 
One may notice from Fig.~\ref{fig:4} that before and after the
collision the boundary lives in flat spacetime. These parts do not
contribute into $S_{tot}$. We find, however, that the collision
point $\tau_\times$ corresponds to a non--analytic
break of the boundary with the extrinsic curvature proportional
to a $\delta$--function,
\begin{equation}\label{extr-curv-boundary} 
K_{\phi=\phi_0}=2\delta(\tau_0-\tau_{0,\times})\left[\mathrm{arcsh}\sqrt{-V_{\mathrm{eff}}}-\mathrm{arcsh}\sqrt{-V_{\mathrm{eff}}/f}\right]\Bigl|_{r=r_0}\;,
\end{equation}
where $\tau_0$ is the boundary proper time and $\tau_{0,\times}$ is
its value at the collision point. The expression
\eqref{extr-curv-boundary} is complex because $f<0$ at
$r=r_0<r_h$. Substituting it into Eq.~\eqref{eq:grav-action}, one
finds an imaginary term 
\begin{equation}\label{imag-action1}
\Im m \,S_{tot}^{(2)}
=\frac{M_{cr}}{\lambda}\,\Im
m\,\log\bigg(1-\frac{M+i\varepsilon'}{M_{cr}}\bigg)
=-\frac{\pi}{\lambda} M_{cr}\,\theta(M-M_{cr})\;.
\end{equation}
There are no imaginary contributions
in addition to Eqs.~\eqref{imag-time} and \eqref{imag-action1}. Summing
up these terms, we arrive to the expression (\ref{imag-action}).

\section{Relation to black  entropy}
\label{sec:suppression}

\subsection{Euclidean calculation of entropy: a puzzle}
\label{sec:dinaive-approach}

Our semiclassical result \eqref{imag-action} is natural from the
quantum-mechanical viewpoint: the probability $\w{P}_{fi}$ of
particle reflection  is a continuous function of energy $M$, as it
should be. At $M>M_{cr}$ i.e.\ above the threshold for classical 
BH production, this probability is exponentially
suppressed. The respective  transitions are interpreted as two-stage
processes. First, the left-moving particle creates the 
BH of mass $M$ classically. Second, the intermediate BH decays into
the final-state 
particle with exponentially small probability. One expects
\cite{Parikh:1999mf, Parikh:2004ih} that the probability of the latter
stage is suppressed by the number of BH states
$\exp(\Sigma_{BH})$. This implies the expression \eqref{corr-entr} for
the black hole entropy $\Sigma_{BH}$.

\begin{figure}[t]
  
\centerline{\includegraphics{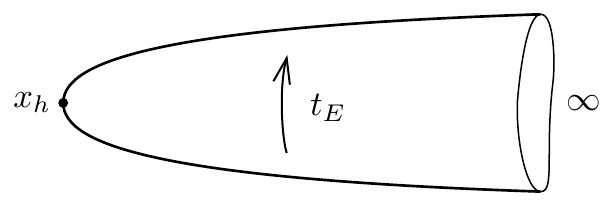}\hspace{1cm}
  \includegraphics{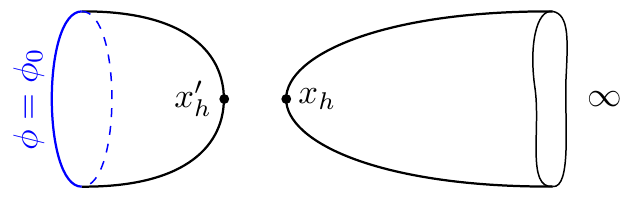}}

\hspace{3.5cm}(a) \hspace{7.5cm}(b)

\caption{Geometries for Euclidean calculations of the black hole
  entropy.} \label{fig:5}
\end{figure}

A following puzzle arises. There is an alternative method for calculating BH
entropy based on Euclidean path integral \cite{Gibbons:1976ue}. When
applied to our model, this method 
apparently gives a different result (\ref{Snaive}). 
Let us briefly review the relevant
calculation~\cite{Hayward:1994dw, Solodukhin:1995te}. 
One computes the thermal partition function 
\begin{equation}\label{th-partition}
\w{Z}(\beta)=\int\limits_{\rm periodic} {\cal D}\Phi\,\e^{-S_E[\Phi]}\;,
\end{equation} 
where $S_E$ is the Euclidean CGHS action, 
see Appendix~\ref{sec:euclidean-entropy} for the precise
  definition. The integral is taken over configurations with period
  $\beta$ in Euclidean time $t_E=it$.
In the semiclassical limit the integral is saturated by the saddle
point. The
instanton
corresponds to Euclidean continuation of the BH exterior with
the metric
\begin{equation}\label{eucl-metric}
ds^2=f(r)\, dt_E^2+\frac{dr^2}{f(r)}\;, \qquad\qquad t_E\in[0;\beta]\;.
\end{equation}
This spacetime has topology of a half-tube, where 
 $t_E$ serves
as a periodic coordinate, see Fig.~\ref{fig:5}a.
The black hole horizon corresponds to the
tip of the tube. Notably, this tip is a conical singularity if the period
$\beta$ is not equal to the inverse Hawking temperature
$T_H^{-1}=2\pi/\lambda$. 
As a consequence, the curvature has a $\delta$-function contribution
at the tip of the cone,
\begin{equation}\label{tube-curv}
R=4\pi(1-\beta T_H)\;\frac{\delta^{(2)}(x-x_h)}{\sqrt{g}}+2\lambda M
\mathrm{e}^{-2\lambda r}\;.
\end{equation}
The singular contribution vanishes for $\beta=T_H^{-1}$.

Now one evaluates the Euclidean action on this solution,
\begin{equation}\label{eucl-action}
S_E=M(\beta-T_H^{-1})\;,
\end{equation}
and the free energy,
\begin{equation}
F(\beta)\equiv -\beta^{-1} \log \w{Z}(\beta)\simeq \beta^{-1}S_E(\beta)\;.
\end{equation}
Note that the only non-vanishing contribution
into the action comes from the $\delta$--function in
Eq.~(\ref{tube-curv}). Then the thermodynamical formula
\begin{equation}
  \label{eq:2}
\Sigma_{BH}=\beta^2\frac{\partial
  F}{\partial\beta}=\beta\frac{\partial S_E}{\partial\beta}-S_E\; 
\end{equation}
yields the ``naive'' entropy (\ref{Snaive}).

Thus, we have two different expressions for the BH entropy ---
Eqs.~(\ref{corr-entr}) and (\ref{Snaive}) --- and we have to decide
which one is correct.\footnote{Note that both expressions agree
  with the first law of BH thermodynamics 
$T_{H}\Delta \Sigma_{BH}=\Delta M$.}

\subsection{Experiments with the thermal gas}
\label{sec:exper-with-therm}

We now present several physical arguments against Eq.~(\ref{Snaive}). 
To this end, we couple the dilaton gravity to
the gas of massless 
particles -- quanta of some massless scalar field. Notice that
the BHs cannot form classically from arbitrary configuration of this
field, even if its total mass is higher than $M_{cr}$. Indeed, the 
gravitational Lagrangian (\ref{eq:grav-action}) is explicitly
proportional to the factor ${\mathrm{e}^{-2\phi}\equiv \e^{2\lambda r}}$. This
means that the gravitational interaction decreases exponentially at
coordinate distance $\Delta r\sim\lambda^{-1}$ from the boundary. Then
 formation of BHs requires the
  energy $M_{cr}$ to be concentrated within the interval $\Delta r\sim
  \lambda^{-1}$. A configuration satisfying this condition,
  however, cannot carry large 
coarse--grained entropy. Indeed, the 
entropy reaches maximum in a thermal state providing the bound 
\begin{equation}\label{gas-entr-bound}
\Sigma_{gas}\leq\frac{2M_{cr}}{T_{gas}}
\sim \e^{-\phi_0}\;.
\end{equation}
Here we related the gas temperature to its energy density
$T_{gas}=\sqrt{6\rho_{gas}/\pi}$ and substituted 
$\rho_{gas}\sim M_{cr}/\Delta r$. 
On the other hand, the entropy \eqref{Snaive} is parametrically
larger than Eq.~(\ref{gas-entr-bound}): 
$\Sigma_{BH}^{naive}=4\pi \e^{-2\phi_0}$ at $M=M_{cr}$.
If this expression were correct, it would be puzzling why the
critical BH cannot
be formed from states with large entropy.

Further, the results on classical subcritical
scattering \cite{Fitkevich:2017izc} suggest that the entropy of the
critical BH is even smaller than (\ref{gas-entr-bound}).
Namely, near the threshold of critical BH formation reflection of
the classical field from the boundary proceeds as follows. 
A part of the incoming wavepacket reflects immediately, 
whereas the remaining part forms
a long-lived state with mass $M\approx M_{cr}$. The latter state decays
into a narrow wavepacket carrying a few highly blue-shifted
particles. This may be interpreted as formation of a slightly
subcritical black hole decaying classically into a low-entropy
state. Extrapolating this picture to the critical black hole, we
conclude that it should have an order-one entropy.
Our semiclassical formula \eqref{corr-entr} is consistent with this
picture.

One may also try to form the black hole in an essentially
quantum way. Namely, suppose the spacetime is filled with a massless
gas of temperature $T\sim\lambda$. Eventually, the black hole of mass
$M$ may appear, eating a part of the gas and providing the first order
phase transition. According to Eq.~\eqref{corr-entr}, a part of gas
entropy ${\Delta\Sigma_{gas}=2M_{cr}/\lambda\sim \e^{-2\phi_0}}$
disappears in this process even if the black hole is critical. We have
argued, however, that the black hole cannot form in classical
  collapse of the
low--temperature gas. Thus, the probability of this
process is exponentially suppressed by the CGHS action $S_{gr}\propto
\e^{-2\phi_0}\sim\Delta \Sigma_{gas}$. Now, we recall that the
entropy of a thermal ensemble can decrease 
with exponentially small probability due to
large fluctuations. The above process appears to be one of them. 

It could appear that the problem with entropy might be fixed by adding
to the Euclidean action 
a topological term
\begin{equation}
  \label{eq:3}
  \Delta S_{E}^{(\chi)} = 4\pi Y \chi = Y \int\limits_{\phi < \phi_0} d^2 x_E \sqrt{g} \, R +
  2Y \!\!\!\int\limits_{\mathrm{boundary}}\!\!\! d\tau_0 \; \kappa K
\end{equation}
with $\kappa =+1$.
Here ${\chi = 2 - 2{\rm g} - b}$ is the Euler characteristic of spacetime
with ${\rm g}$ handles and $b$ boundaries. Being a topological invariant, it
does not affect the 
semiclassical dynamics. Also, the
    action of the flat vacuum with $\chi=0$ remains unchanged.
At the same time, the new term adds
a constant $\Delta S_E^{(\chi)} = 4\pi Y$ to the action of the
instanton in Fig.~\ref{fig:5}a and therefore shifts
the entropy in Eq.~(\ref{eq:2}) by $-4\pi Y$. This reproduces
Eq.~\eqref{corr-entr} if $Y = \mathrm{e}^{-2\phi_0}$. Note, however,
that the term~(\ref{eq:3}) with positive $Y \gg 1$ leads to severe
divergence\footnote{Note that $g_s \sim \mathrm{e}^{4\pi Y}$ is a direct
    analog of the coupling constant in string theory.} of the path
integral~\eqref{th-partition} due to exponentially enhanced
contributions of multihandle geometries. Thus, it introduces strong
coupling and does not cure the problem.

To see this more explicitly, let us focus on the case with massless
matter. Then, at the classical level,
the
term~(\ref{eq:3}) can be completely absorbed by the 
field redefinition
\begin{equation}
  \label{eq:4}
  g_{\mu\nu} = \frac{g_{\mu\nu}'}{1 + Y \mathrm{e}^{2\phi'}} \;, \qquad
  \qquad \mathrm{e}^{-2\phi} = \mathrm{e}^{-2\phi'} + Y\;.
\end{equation}
This gives the CGHS action~(\ref{eq:grav-action}) for $g_{\mu\nu}'$
and $\phi'$ with two different parameters: new semiclassical constant
$\mathrm{e}^{2\phi_0'} = [\mathrm{e}^{-2\phi_0} - Y]^{-1}$ in place of
$\mathrm{e}^{2\phi_0} \ll 1$  and new ``mass'' parameter $\lambda' = \lambda
\mathrm{e}^{\phi_0' - \phi_0}$ in the boundary term\footnote{The 
    bulk parameter $\lambda$ remains
    unchanged. Recall that we related the boundary ``mass'' to
    $\lambda$ by requiring existence of a flat vacuum. This condition
    is not Weyl invariant and therefore broken by
    Eq.~\eqref{eq:4}.}. The choice $Y = \mathrm{e}^{-2\phi_0}$
corresponds to a strongly coupled model with 
$\phi_0' = +\infty$.

\subsection{Correcting the Euclidean calculation}
\label{sec:correct-Euclid}

We now suggest a modification of the 
Euclidean calculation that reproduces the result~\eqref{corr-entr} for
the entropy. 
The
approach of Sec.~\ref{sec:dinaive-approach} misses an important
property of our model, namely, the presence of the boundary at
$\phi=\phi_0$. This boundary is necessary because it 
shields the singularities of the CGHS
fields in the original Lorentzian path integral
\cite{Russo:1992ax, Banks:1992ba, Russo:1992ht, deAlwis:1992emy,
Thorlacius:1994ip, Fitkevich:2020okl}.
Our complex scattering solutions
satisfy this property, whereas the Euclidean instanton in
Fig.~\ref{fig:5}a does not.

We cure this problem by adding to the Euclidean spacetime a
disjoint cap-like portion with a
closed boundary $\phi=\phi_0$, see Fig.~\ref{fig:5}b. 
By Birkhoff theorem, the geometry  of the cap is given by the black
hole metric (\ref{eucl-metric}), possibly with a different mass
parameter $M'$. The latter must be larger than $M_{cr}$ for the cap to
be compact and satisfy the inequality $\phi<\phi_0$. The radial
coordinate on the cap runs in the interval
$r_0<r<r_h(M')$. Importantly, the signature of the metric on the cap
is $(-,-)$ instead of $(+,+)$ in the exterior region.

This configuration does not satisfy the boundary condition
(\ref{dil-bc}) at $\phi=\phi_0$ and thus it is not an exact saddle point
of the path integral (\ref{th-partition}). Rather, as shown in
Appendix~\ref{sec:euclidean-entropy}, it should be interpreted as a
constrained instanton extremizing the Euclidean action within a subset of
geometries with the boundary.
Instead of solving 
the boundary conditions, one minimizes the action
with respect to the free parameter~$M'$.

The action of the additional Euclidean cap equals (see
Appendix~\ref{sec:euclidean-entropy}) 
\begin{equation}\label{excess-entropy} 
\Delta
S_E=M'/T_H\;,
\end{equation}
where the terms proportional to $\beta$ have cancelled. The only
remaining contribution comes from the $\delta$-function
 of the curvature
at the horizon $r_h'$. The latter has an
 opposite sign to that in Eq.~(\ref{tube-curv}) due to the metric
 signature $(-,-)$. 
The minimum of $\Delta S_E$ is reached at the boundary $M' \to
  M_{cr}$ of the parameter region where the solution in
  Fig.~\ref{fig:5}b exists. 

Adding up Eqs.~\eqref{eucl-action} and (\ref{excess-entropy}) at
$M'=M_{cr}$ one reproduces Eq.~\eqref{corr-entr}.
  This restores agreement between the semiclassical entropy
  and the scattering probability.

In the generalized model with topological term (\ref{eq:3}) one still obtains
correct entropy (\ref{corr-entr}). Indeed, the additional cap in
Fig.~\ref{fig:5}b has the same topology as the original
Gibbons-Hawking instanton, but its contribution\footnote{In
  this case $\kappa = n_0^{\mu} n_{0\, \mu} = -1$,   or the term is
  not a topological invariant.} into Eq.~(\ref{eq:3}) has opposite sign due
to $(-,\, -)$ signature. Thus, $\Delta S_{E}^{(\chi)}=0$ for any $Y$.

\section{Conclusions}
\label{sec:conclusion}
In this paper we further developed complex semiclassical method for
calculating ${\cal S}$--matrix elements in gravity. We considered a simplified
setup where the point--like quantum particle scatters off the boundary
in two--dimensional Callan--Giddings--Harvey--Strominger (CGHS)
model. The semiclassical method provided the amplitude of a complete
transition between the initial particle moving with energy $M$ towards
the boundary and an outgoing final particle with the same energy. At
low energies this reflection proceeds classically and the transition
probability is of order one. However, once the particle energy exceeds the
minimal mass of black holes (BHs) in the model, the amplitude becomes
exponentially suppressed. Then the respective transition can be
interpreted as production of an intermediate BH and its
subsequent decay into an outgoing particle. The probability of such
transition is naturally identified with $\exp(-\Sigma_{BH})$,
where $\Sigma_{BH}$ is the BH entropy. It is important to stress that
our analysis provides not only the absolute value of the amplitude,
but also its phase.

Our result implies that the entropy of the minimal-mass BH
vanishes. This is consistent with the expressions for BH
entropy in similar two-dimensional models 
\cite{Fiola:1994ir,Myers:1994sg,Hayward:1994dw,Solodukhin:1995te}. 
We noticed, however, an apparent conflict between this result and the
calculation of the BH entropy using the Gibbons--Hawking
Euclidean approach. We suggested a natural modification of the
Euclidean calculation that takes into account the presence of the
boundary and recovers the correct entropy obtained from the scattering
probability. 

Our results demonstrate that the semiclassical ${\cal S}$-matrix provides
important insights about black holes, even if the simplified matter
content is considered. It is straightforward to apply our approach to
spherically--symmetric sectors of multidimensional
gravities. In particular, the case of $4$ dimensions was
considered in~\cite{Bezrukov:2003tg}.  A
simplified matter content in this case is provided by thin
spherical shells with dynamical radius $R = R(\tau)$. 

Let us outline several directions for future
research. 

The phase of the amplitude is known to contain information about
temporal properties of the scattering
process~\cite{Landauer:1994zz}. It will be interesting to extract this
information from our results and compare it to the characteristic time
scales of the BH evaporation, e.g. the scrambling time
\cite{Hayden:2007cs}.

Further development of the semiclassical ${\cal S}$-matrix
approach will be inclusion of full-fledged matter fields. In field theory,
the semiclassical amplitudes can be used for studying quantum
correlations in the Hawking radiation and for direct tests of
unitarity. As an example, consider the identity satisfied in any
$(d+1)$-dimensional 
unitary theory,
\begin{equation} 
  \label{eq:1}
  \mathrm{e}^{\int d^d\boldsymbol{k} \, a_{\boldsymbol{k}}^*
    b_{\boldsymbol{k}}}  = \langle a | b\rangle = \langle
  a | \hat{\cal S}^\dag \hat{\cal S} | b \rangle = \int
         {\cal D} c^*  {\cal D} c \; \mathrm{e}^{- \int d^d\boldsymbol{k} \, c_{\boldsymbol{k}}^*
           c_{\boldsymbol{k}}} \;
         \left[\langle c| \hat{\cal S} |a \rangle\right]^* \langle c | \hat{\cal S} |b\rangle\;,
\end{equation}
where $|a\rangle$, $|b \rangle$ and $|c\rangle$ are the flat-space
coherent states in the beginning and end of the scattering
process. One can write the r.h.s.\ in Eq.~(\ref{eq:1}) as a path
integral using Eq.~(\ref{grav-path-int}). At large $a_{\boldsymbol{k}}$ and $b_{\boldsymbol{k}}$ the
initial states are semiclassical. If they are different, their overlap
is exponentially suppressed. Then the integral on the r.h.s.~can be 
evaluated in the saddle-point approximation. Importantly,  the relevant
saddle-point solutions should 
interpolate between flat spacetimes in the initial and final
asymptotic regions.
Comparing the saddle--point result to the
l.h.s.\ of Eq.~(\ref{eq:1}), 
one will perform a nontrivial check of unitarity.

In the context of 2-dimensional dilaton gravity one can add one-loop corrections
by including the Polyakov effective action~\cite{Callan:1992rs} and,
optionally, Russo--Susskind--Thorlacius (RST)
counterterm~\cite{Russo:1992ax, Fitkevich:2020okl}. This modification
may clarify relation between our semiclassical ${\cal S}$-matrix and the
conventional black hole evaporation due to one-loop quantum corrections.
Note, however, that the Polyakov term is nonlocal and therefore introduces an
additional effective field into the model. Solving the semiclassical equations in
this case will require full field-theoretical treatment and goes beyond
the scope of this paper. 

Another interesting direction of research would be to relate 
the semiclassical ${\cal S}$-matrix to the new ``island'' method for
calculating the entanglement entropy of the 
Hawking radiation~\cite{Penington:2019npb,Almheiri:2019psf,Almheiri:2019hni,
Gautason:2020tmk,Hartman:2020swn,Almheiri:2019qdq,Penington:2019kki}.
The latter method indicates
purification of the Hawking radiation in the final state. 
Technically, it makes use of 
``replica wormholes'' \cite{Almheiri:2019qdq,
  Penington:2019kki, Faulkner:2013ana, Hubeny:2007xt,
  Engelhardt:2014gca}, saddle points of the gravitational path 
integral for the trace ${\rm Tr}\, \hat{\rho}^n$, where 
$\hat{\rho}$ is the density matrix of the radiation and $n$ is 
an arbitrary power. If BH is formed from a pure state $\ket{\Psi_i}$, 
the final density matrix equals 
$\hat{\rho} = \hat{\cal S}
| \Psi_i\rangle \langle \Psi_i | \hat{\cal S}^\dag$. Thus,
${\rm Tr}\, \hat{\rho}^n$ can be formally written using the path integral
(\ref{grav-path-int}) for the ${\cal S}$-matrix.
This suggests that the relevant saddle--point solutions in the two
methods may be related to each other by some kind of analytic
continuation.

\paragraph*{Acknowledgments} We thank V.~Rubakov, G.~Rubtsov and P.~Tinyakov 
for encouraging
interest.
This work was supported by the grant
RSF 16--12--10494.

\appendix
\section{Classical solutions}
\label{sec:classical-solutions}
In this Appendix we summarize the field equations 
and discuss the relevant solutions.
 
\subsection{Birkhoff theorem}
\label{sec:equations} 

Varying the action \eqref{eq:grav-action}, \eqref{eq:partic-action}
with respect to $g_{\mu\nu}$ and $\phi$, we find,  
\begin{align}
&\nabla_\mu\nabla_\nu\phi+g_{\mu\nu}\left[(\nabla\phi)^2-\Box\phi-\lambda^2\right]=\e^{2\phi}\,
T_{\mu\nu}/4\;, \label{eq:enst_eqs}\\
&(\nabla\phi)^2-\Box\phi-\lambda^2=R/4\;, \label{eq:diltn_eq}
\end{align}
where $\Box\equiv\nabla_\mu\nabla^\mu$ is the covariant d'Alembertian and
\begin{equation}\label{eq:iv}
T_{\mu\nu}=-\frac{2}{\sqrt{-g}}\,\frac{\delta S_m}{\delta g^{\mu\nu}}
\end{equation}
is the matter energy--momentum tensor. It will be discussed later.

In an empty spacetime region one sets $T_{\mu\nu}=0$ and arrives to the system
\begin{equation}\label{eq:vacuum}
2\nabla_\mu\nabla_\nu\phi=g_{\mu\nu}\Box\phi\;, \qquad
(\nabla\phi)^2-\frac12\Box\phi=\lambda^2\;, \qquad \Box\phi=-R/2\;,
\end{equation}
where  the second equation is a trace of Eq.~(\ref{eq:enst_eqs}).
It will be convenient to use the Schwarzschild gauge,
\begin{equation}\label{eq:schw-gauge}
ds^2=-h(r,t)\,dt^2+\frac{dr^2}{f(r,t)}\;,~~ \qquad \phi=-\lambda r\;,
\end{equation}
where the spatial coordinate $r$ tracks the dilaton field $\phi$ and
the time $t$ is orthogonal to $r$. The first of Eqs.~\eqref{eq:vacuum}
gives, 
$$ \partial_t f = \partial_r (h/f)=0\;, $$ 
implying that the metric component $f=f(r)$ is time--independent and
${h=c(t)f(r)}$ with arbitrary $c(t)$. One can fix $h=f(r)$
using the residual time reparametrization invariance in
Eq.~\eqref{eq:schw-gauge}. 

With these simplifications the second of Eqs.~\eqref{eq:vacuum} reduces to
\begin{equation}
\partial_r f = 2\lambda(1-f)\;,
\end{equation}
with general solution \eqref{sol:black}. The Schwarzschild mass $M$
is an arbitrary integration constant in this solution. 
It is straightforward to check that the third of
Eqs.~\eqref{eq:vacuum} is now automatically satisfied.  

To summarize, we demonstrated that the static black hole spacetime
\eqref{sol:black}
with arbitrary mass $M$ is the only solution in an empty patch of
spacetime. This is the analog of the Birkhoff theorem in the present
context \cite{LouisMartinez:1993cc}.

\subsection{Junction conditions and equation of motion for the particle}
\label{sec:junction}
It will be convenient to introduce Gaussian normal coordinates
$(\tau,n)$ near the particle trajectory $x^\mu_\ast(\tau)$. Here $n$
measures the geodesic distance to the trajectory and $\tau$ is orthogonal
to $n$, 
\begin{equation}\label{eq:shell_gauge}
ds^2=-a(\tau,n)\, d\tau^2+dn^2\;.
\end{equation}
In these coordinates the particle trajectory is $n=0$. We choose $\tau$
to coincide with the proper time along the trajectory:
$a(\tau,0)=1$. By construction,
$a(\tau,n)$ is continuous at $n=0$, as opposed to the metric
components in the Schwarzschild gauge. 

Variation of the action \eqref{eq:partic-action} with respect to
$g^{\mu\nu}$ gives the particle energy--momentum tensor \eqref{eq:iv}, 
$$ T^{\mu\nu}=m\,\dot{x}_{*}^\mu\dot{x}_*^\nu\, \delta(n)\;, $$
where $(\dot{x}_*^\tau,\dot{x}_*^n)=(1,0)$ is the particle
velocity. Since Eq.~\eqref{eq:enst_eqs} has a $\delta$--function in
the r.h.s., the normal derivatives of $a$ and $\phi$ are
discontinuous at $n=0$. Equations \eqref{eq:enst_eqs},
\eqref{eq:diltn_eq} take the from 
\begin{align}
&\partial_n^2\phi=m\,\e^{2\phi}\delta(n)/4+(\text{regular terms})\;, \notag\\
&\partial_n(\partial_n a/a)=4\partial_n^2\phi+(\text{regular terms})\;, \label{eq:singular}
\end{align}
where we have kept only the ``singular'' terms with the second
$n$--derivatives, which are proportional to $\delta(n)$. Now, we
integrate Eqs.~\eqref{eq:singular} from $n=-0$ to $n=+0$ and rewrite
them in the covariant form using $\partial_n\phi=n^\mu\nabla_\mu\phi$
and $\partial_n a/(2a)=K$. We arrive to the junction conditions
\eqref{israel}. 

Since $n^\mu\nabla_\mu\phi$ and $K$ are frame--independent, one can
compute them in different coordinate systems $(T,r)$ and $(t,r)$ at the
two sides of the particle trajectory. The outer trajectory normal in
these regions is 
\begin{equation}\label{outer-normals}
(n^T,n^r)=(\dot{r}_\ast,\,\dot{T}_\ast)\;, \qquad\qquad (n^t,n^r)=\left(\frac{\dot{
r}_\ast}{f},\;\dot{t}_\ast f\right)\;,
\end{equation}
where $\dot T_\ast$ and $\dot t_\ast$ 
can be expressed from Eqs.~\eqref{sol:flat} and \eqref{sol:black},
$$ \dot{T}_\ast=\sqrt{1+\dot{r}_\ast^2}\;, \qquad\qquad \dot{t}_*=\frac{\sqrt{f(r_\ast)+\dot{r}_\ast^2}}{f(r_\ast)}\;. $$
Substituting the normal into the first of Eqs.~\eqref{israel} 
one obtains the energy conservation law
\begin{equation}\label{eq:particle-eq-app}
M=m\sqrt{1+\dot{r}_*^2}-\frac{m^2}{8\lambda}\e^{-2\lambda r_*}\;.
\end{equation}
The second junction condition in Eqs.~\eqref{israel} is a time
derivative of Eq.~(\ref{eq:particle-eq-app}). It is trivially
satisfied once Eq.~\eqref{eq:particle-eq-app} is solved. 
Equation of motion \eqref{eq:effective_particle} from the main text is
obtained by squaring Eq.~\eqref{eq:particle-eq-app}. 

\subsection{Boundary condition and  reflection law}
\label{sec:refl-laws}
The saddle--point configurations $\Phi_{cl}(x)$ should extremize the
action \eqref{eq:grav-action} with respect to all variables, in
particular, the metrics $g_{\mu\nu}$ at the boundary
$\phi=\phi_0$. Due to the reparametrization invariance, it is enough to
consider only the variations preserving the coordinate position of
this boundary. Then $\delta\phi=0$ and $\delta n_{0\,\mu}\propto n_{0\,\mu}$ at
the line $\phi=\phi_0$. We vary Eq.~\eqref{eq:grav-action} with
respect to $g_{\mu\nu}$ and leave only the boundary terms,
$$ \delta S_{gr}=2\e^{-2\phi_0}\int\limits_{\phi=\phi_0}d\tau_0\,\left(n_0^\kappa\nabla_\kappa\phi-\lambda\right)\,\tau^\mu\tau^\nu\,\delta g_{\mu\nu}\;, $$
where $\tau^\mu=dx^\mu/d\tau_0$ is the unit vector along the line
$\phi=\phi_0$. Requiring the variation to vanish, we obtain Eq.~\eqref{dil-bc}. 

To derive reflection law for the particle from the boundary, 
we notice that the collision point
$\tau_\times$ divides the particle trajectory $x^\mu_\ast(\tau)$ into
two smooth  parts, see Fig.~\ref{fig:4}. Thus,
\begin{equation}\label{refl-law-deriv}
S_m=-m\int\limits_{\tau_i}^{\tau_\times}d\tau-m\int\limits_{\tau_\times}^{\tau_f}d\tau\;,
\end{equation}
where $\tau_i$ and $\tau_f$ are the initial and final times of the
process. We vary Eq.~\eqref{refl-law-deriv} with respect to $x_*^\mu(\tau)$,
again keeping the position of the boundary intact: $n_{0\, \mu} \delta 
x_*^\mu(\tau_\times)=0$. We obtain,
$$ \delta
S_m= m\tau_\nu\delta x_*^\nu
\left[\tau_\mu\,\dot{x}_*^\mu(\tau_\times+0)
    -\tau_\mu\,\dot{x}_*^\mu(\tau_\times-0)\right]\;, $$
where again only the boundary terms are shown. We obtain two equations,
\begin{align}
\tau_\mu\,\dot{x}_*^\mu(\tau_\times-0)=\tau_\mu\,\dot{x}_*^\mu(\tau_\times+0)
\;,\qquad\quad
n_{0\,\mu}\,\dot{x}_*^\mu(\tau_\times-0)=-n_{0\,\mu}\,\dot{x}_*^\mu(\tau_\times+0)\;, 
\label{eq:part-ref}
\end{align}
where the second one follows from the normalization 
$\dot{x}_*^\mu\,\dot{x}_{*\,\mu}=-1$.

Now, we rewrite Eqs.~\eqref{eq:part-ref} in the coordinates $(T,r)$ of
flat spacetime patches immediately prior to the collision and after it,
see Fig.~\ref{fig:4}. This gives the reflection law 
\begin{equation}
\dot{T}_*(\tau_\times-0)=\dot{T}_*(\tau_\times+0)\;, \qquad\qquad
\dot{r}_*(\tau_\times-0)=-\dot{r}_*(\tau_\times+0)\;, 
\end{equation}
which is used in the main text. 

\section{Regularization method}
\label{sec:regul-method}
Let us demonstrate that the regularization \eqref{eq:int-time},
\eqref{reg-action} of the classical action is equivalent to the
imaginary shift of the Schwarzschild mass $M$ inside the regulating
``shell''
$r_\ast<r<r_\varepsilon$, where
$r_\varepsilon=-\phi_\varepsilon/\lambda$. 
To this end, we solve the field equations at
$r>r_\ast$. Additional term in the regularized action
\eqref{reg-action} produces 
imaginary energy--momentum tensor, 
$$
T_{\mu\nu,\, \varepsilon}=i\varepsilon\,L(\phi)\,(\lambda^2-(\nabla\phi)^2)\,
\big[4\nabla_\mu\phi\nabla_\nu\phi+g_{\mu\nu}(\lambda^2-(\nabla\phi)^2)\big]\;, $$
in the right--hand side of Eq.~\eqref{eq:enst_eqs}. 
In the Schwarzschild gauge (\ref{eq:schw-gauge}) 
the $(rt)$ and $(tt)$ components of Eq.~\eqref{eq:enst_eqs} give 
\begin{align}
\partial_t f=0\;,\qquad\quad
\partial_r\tilde{M}^{-1}(r)=\frac{i\varepsilon}{4}\lambda^2L(-\lambda r)\,\e^{-4\lambda r}\;, \label{regul-mass}
\end{align}
where $\tilde{M}(r)$ is the coordinate-dependent mass entering the metric as
$$ f(r)=1-\frac{\tilde{M}(r)}{2\lambda}\,\e^{-2\lambda r}\;. $$
Integrating the second of Eqs.~\eqref{regul-mass} one arrives to the
matching condition 
\begin{equation}\label{sewing}
\frac1{M}-\frac1{M_\varepsilon}=\frac{i\varepsilon\lambda^2}{4}\int
dr\,L(-\lambda r)\, \e^{-4\lambda r}=\frac{i\varepsilon}{4\lambda}\;.
\end{equation}
Here $M=\tilde{M}(+\infty)$ is the real conserved 
energy of the particle and
${M_\varepsilon=\tilde{M}(r<r_\varepsilon)}$ is the Schwarzschild mass
parameter inside the regulating ``shell'' at
$\phi=\phi_\varepsilon$. Expressing $M_\varepsilon$ from Eq.~\eqref{sewing}, one obtains Eq.~\eqref{complex-mass} from the main text.

\section{Computing the action}
\label{sec:grav-act}
Let us calculate the total action \eqref{eq:total} on the
semiclassical solution. We work in the limit $\varepsilon\to +0$.
To start, we note that the bulk
Lagrangian in Eq.~\eqref{eq:grav-action} is a total derivative on the
field equation~\eqref{eq:diltn_eq}, 
\[
 \e^{-2\phi}\left[R+4(\nabla\phi)^2+4\lambda^2\right]=2\,\Box
 \e^{-2\phi}\;. 
\]
Thus, the interacting action (\ref{eq:action-sum}) 
is a sum of one--dimensional contour
integrals over the  spatial infinity $r=r_\infty\to+\infty$,
the boundary $\phi=\phi_0$, Cauchy surfaces
$t=t_i,t_f$, and the particle
worldline $r=r_\ast(\tau)$, see Fig.~\ref{fig:4}, 
\begin{equation}
\label{Sintterms}
S(t_f,t_i)=S_{r_\infty}+S_{\phi=\phi_0}+S_{t_i}+S_{t_f}+S_m\;.
\end{equation}
This expression includes the Gibbons--Hawking term,
\begin{equation}\label{eq:gibbons}
S_{GH}=2\int\limits_{\infty}d\sigma\,\e^{-2\phi}\kappa\, (K-K_0)\;,
\end{equation}
over the spatial infinity $r=r_\infty$ and surfaces
$t=t_i,t_f\to\mp\infty$. Here $\sigma$ is the proper time or
the proper distance, $K$ is the outer--normal extrinsic curvature, and
$\kappa=n_\mu n^\mu$ equals $+1$ ($-1$) at the timelike line
$r=r_\infty$  (spacelike curves $t=t_i,t_f$). The
parameter $K_0$ is introduced in Eq.~\eqref{eq:gibbons} to subtract
the vacuum contribution;
it equals $2\lambda$ at $r=r_\infty$ and zero at $t=t_i,t_f$. 
Recall that in addition to the terms in 
Eq.~(\ref{Sintterms}), the total action
$S_{tot}$ includes the free actions
$S_0$ and the wave functionals $\Psi_{i,f}$.
We now evaluate all the listed contributions one by one.

{\bf Spatial infinity.} It is straightforward to check that the 
term at $r=r_\infty$ vanishes as $\e^{-2\lambda r_\infty}$,
\begin{equation}\label{contr-spatial}
S_{r_\infty}=2\int\limits_{r=r_\infty}d\sigma\,\e^{-2\phi}\left(K-2\lambda-2n^\mu\nabla_\mu\phi\right)\to 0\;,
\end{equation}
where we evaluated $n^\mu\nabla_\mu\phi=-\lambda\sqrt{f}$ and
$K=f'/2\sqrt{f}$ using the metric \eqref{sol:black}, then sent
$r_\infty\to+\infty$. 

{\bf The boundary.} Adding the boundary term in
Eq.~\eqref{eq:grav-action} to the contribution from the bulk action, one
obtains, 
\begin{equation}\label{boundary-contrib}
S_{\phi=\phi_0}=2\,\e^{-2\phi_0}\int\limits_{\phi=\phi_0}d\tau_0\,K\;,
\end{equation}
where the term proportional to $\lambda$ has cancelled due to the 
boundary condition \eqref{dil-bc}. 
Note that
$\phi=\phi_0$ is a straight line in flat spacetime prior to
and after
the collision, see Fig.~\ref{fig:4}. In these regions
$K=0$. Thus, the only non--zero contribution into $S_{\phi=\phi_0}$
comes from the singularity of the extrinsic curvature at the collision
point. 

\begin{figure}[t]
  \centerline{
  \includegraphics{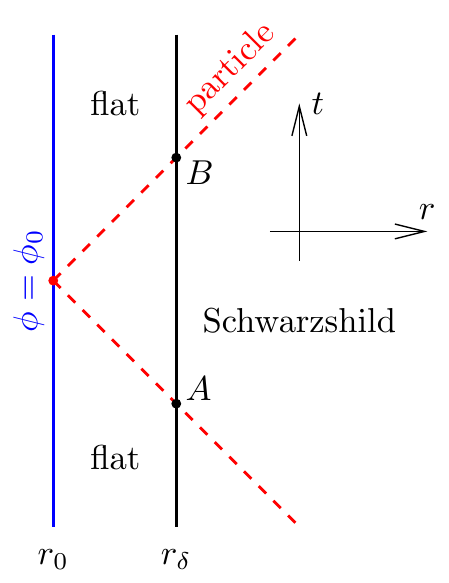}
  \hspace{2cm}
  \includegraphics{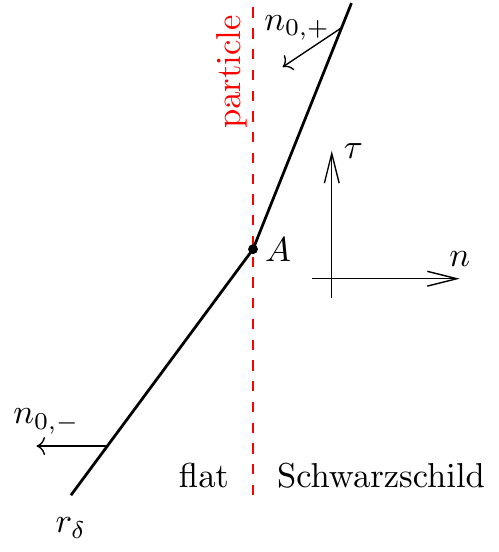}
  }

  \hspace{3.5cm}(a) \hspace{7cm} (b) 

\caption{Regularized boundary $r=r_\delta$ in (a) the original coordinates
   and (b) Gaussian normal frame attached to the particle. 
} \label{fig:9}
\end{figure}

To evaluate it, we use several technical steps. We regulate the
calculation by slightly shifting the line of integration to 
 $r_\delta\equiv r_0+\delta r$, see Fig.~\ref{fig:9}a. 
In
contrast to the boundary, the regulating line intersects the particle
trajectory twice, going from the flat geometry to Schwarzschild and
back. We will see that each of these intersection points gives a 
$\delta$-functional contribution into the integral
(\ref{boundary-contrib}).

Let us focus on the
first intersection point $A$. 
In its vicinity  
we introduce the Gaussian normal coordinates $(\tau,n)$
which are continuous at the particle worldline.
In these coordinates the line $r=r_\delta $ has
a break at $A$, see Fig.~\ref{fig:9}b.
This is because the normal $n_0^\mu$ to this line has a discontinuity,
as we now demonstrate. In the Schwarzschild and flat patches it has
the components 
$$ (n_{0,+}^t,n_{0,+}^r)=\left(0,-\sqrt{f(r_\delta)}\right)\;, 
\qquad\qquad (n_{0,-}^T,n_{0,-}^r)=\left(0,-1\right)\;. $$
At the intersection point $A$ we can decompose $n_{0,\pm}^\mu$ in the
basis of the tangential and normal vectors to the particle
trajectory. In the two patches the former equals to
$$(\tau_+^t,\tau_+^r)=(\dot t_*, \dot r_*)\;,\qquad\qquad
(\tau^T_-,\tau^r_-)=(\dot T_*, \dot r_*)\;,$$
whereas the latter is given by Eq.~(\ref{outer-normals}). In this way we find
the components of $n_{0}^\mu$
in the Gaussian normal frame which are different on the two sides of
the intersection point $A$,
$$(n_{0,\pm}^\tau,n_{0,\pm}^n)=(-\sh\psi_\pm,-\ch\psi_\pm)\;,
\qquad~~ \text{where}~~\sh\psi_+=-\dot{r}_*/\sqrt{f(r_*)}\;,~~~~~
\sh\psi_-=-\dot{r}_*\;.$$ Here
all the quantities are evaluated at $A$. 

Now we regularize the break approximating the
line $r=r_\delta$ with a smooth curve. Its  
normal is 
\begin{equation}\label{regul-normal}
(n^\tau_0,n^n_0)=(-\sh\psi(\tau),-\ch\psi(\tau))\;,
\end{equation}
where 
$\psi(\tau)$ interpolates between $\psi_-$ and $\psi_+$.
The proper time and extrinsic curvature of the curve are
readily computed: $d\tau_0=d\tau/\ch\psi$, 
$K=-\ch\psi\,\partial_{\tau}\psi$. Integrating $K$ in the
vicinity of the point $A$, we obtain,
\begin{equation}\label{calculated-K}
\int\limits_Ad\tau_0\,K=\psi_--\psi_+\;.
\end{equation}
We see that $K$ contains a $\delta$-function at $A$. Another
$\delta$-function with the same coefficient comes from the second
intersection point $B$. 
Taking the limit $r_\delta\to r_0$, we conclude that the 
boundary extrinsic curvature is proportional to
$\delta(\tau_0-\tau_{0,\times})$. Then Eqs.~(\ref{calculated-K}) and 
\eqref{eq:effective_particle} yield
Eq.~\eqref{extr-curv-boundary} from the main text. 

Finally, using the formula for $V_{\rm eff}$ and expressing $r_0$
in terms of $M_{cr}$, we arrive to the boundary
action,
\begin{equation}\label{contr-boundary}
S_{\phi=\phi_0}=\frac{M_{cr}}{\lambda}\log\left(1-\frac{M+i\varepsilon'}{M_{cr}}\right)+\frac{2M_{cr}}{\lambda}\log\left(\frac{4M_{cr}(p_\times+M)+m^2}{4M_{cr}(p_\times+M)-m^2}\right)\;,
\end{equation}
where $p_\times$ is defined in Eq.~(\ref{pcross}) and the imaginary
part of the logarithm is fixed by the regularization procedure from
Appendix~\ref{sec:regul-method}. 

\begin{figure}[t]
  \centerline{
  \includegraphics{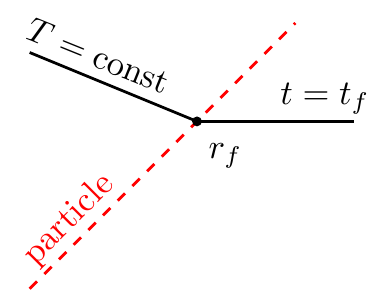}
  }
\caption{The final Cauchy surface.} \label{fig:10}
\end{figure}

{\bf Initial and final Cauchy surfaces.} 
We define them 
as the lines of constant Schwarzschild time $t=t_{i,f}$ to the right
of the initial and final particle positions $r_{i,f}$ 
continued as $T=\mathrm{const}$
to the left, see Fig.~\ref{fig:10}. 
The interacting action at the final surface equals
\begin{equation}
S_{t_f}=-2\int\limits_{t=t_f}d\sigma\,\e^{-2\phi}K\;.
\end{equation}
One can check that $K=0$ on the outer and inner parts of this
surface. Thus, the only non--zero contribution comes
from the jump of the normal at the particle position $r=r_f$ where the
two
parts of the surface join. The same calculation as before gives,\ 
\begin{equation}\label{contr-tf}
S_{t_f}=-2\,\e^{2\lambda r_f}\left[{\rm arcsh}\,\dot r_*-{\rm
    arcsh}\,\frac{\dot r_*}{\sqrt{f}}\right]_{r_f}=\frac{p}{2\lambda}\;,
\end{equation} 
where $p=\sqrt{M^2-m^2}$ and in the second equality we have sent
$r_f\to+\infty$.

Computing the initial contribution at $t=t_i$ in a similar way, one obtains,
\begin{equation}\label{contr-ti}
S_{t_i}=-2\int\limits_{t=t_i}d\sigma\,\ e^{-2\phi}K=\frac{p}{2\lambda}\;,
\end{equation}
which doubles the contribution \eqref{contr-tf}.

{\bf Particle worldline.} The particle action 
 \eqref{eq:partic-action} is already expressed as a contour
 integral. We divide it into two parts, prior to the collision with the
 boundary and after it, 
\begin{equation}\label{eq:part-act-closed}
S_m=-m\int_{r_0}^{r_i}\frac{dr}{\sqrt{-V_{\mathrm{eff}}(r)}}-m\int_{r_0}^{r_f}\frac{dr}{\sqrt{-V_{\mathrm{eff}}(r)}}\;,
\end{equation}
where we also changed the integration variable to $r\in\w{C}_r$ using
Eq.~\eqref{eq:effective_particle}, see Figs.~\ref{fig:contours}a,c, 
and recalled that
reflection flips the sign of $\dot{r}_\ast$. As before, $r_{i,f}$ are
the particle positions at $t=t_{i,f}$. Explicitly
calculating the integral \eqref{eq:part-act-closed} we find, 
\begin{gather}
S_m=\frac{m^2}{\lambda p}\log\,\left(\frac12+\frac{Mm^2}{8M_{cr}p^2}
+\frac{p_\times}{2p}\right)-\frac{m^2}{p}(r_i+r_f-2r_0)\;,
\label{eq:part-act-final}
\end{gather}
where we extracted the asymptotics at $r_{i,f}\to+\infty$. 
Note that this contribution diverges linearly. The
divergence will cancel, however, when we add
the initial and final terms. 

{\bf Initial and final terms.} The expression~\eqref{eq:total} for
$S_{tot}$ 
includes contributions from the initial and final wavefunctions 
$\Psi_{i,f}(r_{\mp})=\exp(\mp ipr_\mp)$, as well as the free actions $S_0$. The latter describe freely moving particle with momenta $\mp p$,
$$ S_0(t_i,0_-)=p(r_--r_i)-Mt_i\;, 
\qquad\qquad S_0(0_+,t_f)=p(r_+-r_f)+Mt_f\;, $$
where $r_\mp$ are the positions of the free particle at $t=0_{\mp}$.
Combining the terms, one obtains 
\begin{equation}\label{contr-free}
S_0(t_i,0_-)+S_0(0_+,t_f)-i\log\Psi_f^\ast-i\log\Psi_i=-p(r_i+r_f)+M(t_f-t_i)\;.
\end{equation}
The change of the 
Schwarzschild time appearing here 
is given by
the 
integral \eqref{time-cont}. Taking it explicitly, we obtain, 
\begin{align}
&M(t_f-t_i)
=
-\frac{M}{\lambda}\log\left(1-\frac{M+i\varepsilon'}{M_{cr}}\right)
-\frac{M^2}{\lambda
  p}\log\left(\frac12+\frac{Mm^2}{8M_{cr}p^2}+\frac{p_\times}{2p}\right)
\notag\\ 
&+ \frac{M}{\lambda}\log\left[
\frac{4M^3-3m^2M+(4M^2-m^2)p_\times}{(M+p)^3}+
\frac{m^2(4M^2+m^2)}{4M_{cr}(M+p)^3}
\right]\!+\!\frac{M^2(r_f+r_i-2r_0)}{p}
\;.\label{Mdt}
\end{align}
Note that the contribution \eqref{contr-free}, \eqref{Mdt} also
diverges as $r_{i,f}\to+\infty$. 
\\

Collecting the terms 
\eqref{contr-spatial}, \eqref{contr-boundary}, \eqref{contr-tf},
\eqref{contr-ti}, \eqref{eq:part-act-final}, \eqref{contr-free}, and
\eqref{Mdt}, one
finally arrives to the total action~\eqref{fully-total}. Note that
the divergences at $r_{i,f}\to+\infty$ cancel between
Eqs.~(\ref{eq:part-act-final}) and \eqref{contr-free}, \eqref{Mdt}. 

\section{Constrained instantons for the entropy}
\label{sec:euclidean-entropy}
In this Appendix we give details of the Euclidean derivation of BH
entropy. Performing the Wick rotation $t=-it_E$ in
Eq.~\eqref{eq:grav-action}, one obtains the Euclidean action, 
\begin{align}\label{eucl-grav-act}
S_{gr,\,E} =  -iS_{gr}= & -\int d^2x_E\,\sqrt{g}\,\e^{-2\phi}\left[R+4(\nabla\phi)^2+4\lambda^2\right]-2\int\limits_{\phi=\phi_0}d\tau_0\,\e^{-2\phi}(\kappa\,K+2\lambda) \notag\\
& -2\int\limits_{r=r_\infty}d\sigma\,\e^{-2\phi}(\kappa\, K-2\lambda)\;,
\end{align}
where we explicitly added the Gibbons--Hawking term at
infinity. The parameter $\kappa=n^\mu n_\mu=\pm 1$ discriminates between
the signatures $(+,+)$ and $(-,-)$ of the Euclidean
spacetime. Note that this parameter is implicitly present\footnote{We
  fixed $\kappa=+1$ in Eq.~\eqref{eq:grav-action} because the
    scattering solutions included timelike boundary.} in the original
Minkowski action, or the latter would be inconsistent.

To warm up, consider the standard Gibbons--Hawking instanton in
Fig.~\ref{fig:5}a. Since the solution \eqref{eucl-metric} is
stationary, one may naively expect that its Euclidean action is
proportional to $\int dt_E=\beta$. This would give zero entropy in
Eq.~\eqref{eq:2}. However, in the vicinity of the horizon
$r-r_h\ll r_h$ the metric \eqref{eucl-metric} takes the form 
\begin{equation}\label{polar}
ds^2=d\rho^2+\rho^2 d\theta^2\;,
\end{equation}
where $\rho=\sqrt{2(r-r_h)/\lambda}$ and $\theta=\lambda t_E$ are the
radial and angular coordinates. Since $\theta$ changes between $0$ and
$\lambda\beta$, this metric describes a cone with angle deficit
$2\pi-\lambda\beta$. The respective $\delta$--contribution in
curvature, Eq.~\eqref{tube-curv}, is proportional to the angle
deficit, not to $\beta$. That is why the standard calculation gives
non--zero black hole entropy. 

Now, let us ensure that every single configuration in the Euclidean
path integral (\ref{th-partition})
includes a boundary $x^\mu=x^\mu_{b}(\tau_0)$ and $\phi$ equals
$\phi_0$ at this boundary. The latter condition is enforced by a
$\delta$--function in the integration measure,
\begin{equation}\label{delta-path-int} 
\prod_{\tau_0}\delta\big(\phi(x_{b}(\tau_0))-\phi_0\big)=\int\w{D}\Lambda\,\e^{-\int
    d\tau_0\,\Lambda(\tau_0)\,[\phi(x_{b})-\phi_0]}\;,
\end{equation}
with the boundary function $\Lambda(\tau_0)$ playing the role of a
Lagrange multiplier. The product on the l.h.s.~is taken over all
points on the boundary.
This adds an extra term to the Euclidean action,
\begin{equation}\label{lambda}
S_E = S_{gr,\,E} + \int
  d\tau_0\;\Lambda(\tau_0)\,(\phi_{b}-\phi_0)\;,
\end{equation}
where $\phi_{b}\equiv\phi(x_{b}(\tau_0))$.

Importantly, the term \eqref{lambda} changes the boundary conditions
at $x=x_{b}$. Indeed, variations with respect to
$g_{\mu\nu}$ and $\phi$ now give\footnote{In general, the second
  of Eqs.~\eqref{bc-dil-mod} is obtained from the first by taking a
  derivative  along the boundary and dividing
  the equation by $d\phi_{b}/d\tau_0$. 
The two equations are independent, however, if $\phi_{b}$ is
constant at $x=x_{b}(\tau_0)$.} 
  \begin{equation}
    \label{bc-dil-mod} 
  n_0^\mu\nabla_\mu\phi=\kappa\lambda-\frac{\kappa}{4} \,
  \e^{2\phi}\Lambda\,(\phi_{b}-\phi_0)\;, \qquad
-2n_0^\mu\nabla_\mu\phi+K+2\kappa\lambda=-\frac{\kappa}{4} \,\e^{2\phi}\Lambda 
\end{equation}
at $x=x_{b}(\tau_0)$. Besides, variation with respect to
$\Lambda(\tau_0)$ gives equation
$\phi_{b}=\phi_0$. In what follows we find solutions at a
fixed $\Lambda$ and then take the limit $\phi_{b}\to\phi_0$.

It is clear that the Gibbons--Hawking instanton in Fig.~\ref{fig:5}a
does not satisfy the  condition $\phi_b = \phi_0$, as it has
$\phi<\phi_h<\phi_0$. Thus, we have to suggest an alternative. Let us
assume that the true saddle--point configuration exists and it is real.
Besides,
we will consider only the solutions with $\tau_0$--independent
$\phi_{b}=-\lambda r_{b}$. This is reasonable because we will eventually
send $\phi_{b}\to\phi_0$. 

With the above assumptions, the only candidate for correct instanton
is the disconnected configuration in Fig.~\ref{fig:5}b. The
Birkhoff theorem guarantees that the additional cap--like part in this
configuration is described by the Schwarzschild metric
\eqref{eucl-metric} with mass parameter $M'$ in place of
$M$. Besides, the instanton should include precisely one
infinity with fixed ADM mass $M$. This specifies the patch
$r_{b}<r<r_h'$ of the cap. Note that the metric
\eqref{eucl-metric} is this case has signature $(-,-)$ and $\kappa=-1$
in Eqs.~\eqref{bc-dil-mod}. 

Substituting Eq.~\eqref{eucl-metric} and constant $\phi_{b}$
into Eqs.~\eqref{bc-dil-mod}, one finds boundary conditions, 
\begin{subequations}
\label{cond-6}
\begin{align}
  &\Lambda(\phi_{b}-\phi_0)=2M_{b}\left(1 + \sqrt{M'/M_{b}-1}
  \right)\;, \\
  & \left(\Lambda/2  + 2M_b\right)\sqrt{M'/M_{b}-1} = 2M_b - M'\;,
\label{cond-6b}
\end{align}
\end{subequations}
where $M_b = 2\lambda \mathrm{e}^{-2\phi_b}$. 
The limit $\phi_{b}\to\phi_0$ corresponds to $\Lambda\to\infty$ and $M'\to
M_{cr}$, with the combinations on the l.h.s. of
Eqs.~\eqref{cond-6} held fixed. 
At this point we obtained a unique solution. 

The only
additional contribution into the Euclidean action which is not
proportional to $\beta$ comes from the conical singularity at its second
horizon $r_h'$. This time, however, the metric in the vicinity
of $r_h'$ is negative--definite: $ds^2=-d\rho^2-\rho^2 d\theta^2$, where $\rho=\sqrt{2(r_h'-r)/\lambda}$ and $\theta=\lambda t_E$. Thus, the respective $\delta$--term of the curvature has opposite sign,
\begin{equation}\label{cap-curva}
R=-4\pi(1-\beta T_H)\,\frac{\delta^{(2)}(x-x_h')}{\sqrt{g}}+2\lambda M' \e^{-2\lambda r}\;,
\end{equation}
cf. Eq.~\eqref{tube-curv}.
Substituting Eqs.~(\ref{cond-6}), (\ref{cap-curva}) into 
Eq.~(\ref{lambda}) one finds the action  
(\ref{excess-entropy}) of the additional cap. The latter is
$\beta$-independent. 
Note that $M'$ in this expression is related to
$\phi_b$ via Eqs.~(\ref{cond-6}). 
The limit $\phi_b \to \phi_0$ implies $M' \to M_{cr}$. 
Then Eq.~\eqref{eq:2} 
reproduces the
entropy \eqref{corr-entr}.

Note that alternatively one can ignore some saddle-point equations and directly
minimize the Euclidean action over the free parameters of the
solutions. In particular, the discussion of
Sec.~\ref{sec:correct-Euclid} corresponds to neglecting
Eq.~(\ref{cond-6b}) and minimizing with respect to the single
remaining parameter $M'$.



\begin{thebibliography}{99}
  \bibitem{Hawking:1976ra}
   S.~Hawking, {\it Breakdown of Predictability in Gravitational Collapse,}
   \href{https://journals.aps.org/prd/abstract/10.1103/PhysRevD.14.2460}{Phys. Rev. D
   \textbf{14} (1976) 2460}.
\bibitem{Harlow:2014yka}
  D.~Harlow, {\it Jerusalem Lectures on Black Holes and Quantum Information,}
  \href{https://doi.org/10.1103/RevModPhys.88.015002}{Rev.\ Mod.\ Phys.\  {\bf
      88} (2016) 015002} [\href{https://arxiv.org/abs/1409.1231}{1409.1231}].
\bibitem{Hawking:1974sw}
  S.~W.~Hawking,
  {\it Particle Creation by Black Holes},
  \href{https://doi.org/10.1007/BF02345020}{Commun.\ Math.\ Phys.\  {\bf
      43} (1975) 199} [\href{https://doi.org/10.1007/BF01608497}{{\it
        erratum-ibid} {\bf 46} (1976) 206}].
\bibitem{Penington:2019npb}
  G.~Penington, {\it Entanglement Wedge Reconstruction and the
    Information Paradox,}
  \href{https://arxiv.org/abs/1905.08255}{1905.08255}.
\bibitem{Almheiri:2019psf}
  A.~Almheiri, N.~Engelhardt, D.~Marolf and H.~Maxfield, {\it The
    entropy of bulk quantum fields and the entanglement wedge of an
    evaporating black hole},
  \href{https://doi.org/10.1007/JHEP12(2019)063}{JHEP \textbf{12}
    (2019) 063}
  [\href{https://arxiv.org/abs/1905.08762}{1905.08762}]. 
\bibitem{Page:1993wv}
  D.~N.~Page, {\it Information in black hole radiation},
  \href{https://doi.org/10.1103/PhysRevLett.71.3743}{Phys. Rev. Lett. \textbf{71}
    (1993) 3743} [\href{https://arxiv.org/abs/hep-th/9306083}{hep-th/9306083}]. 
\bibitem{Hayden:2007cs}
  P.~Hayden and J.~Preskill, {\it Black holes as mirrors: Quantum
    information in random subsystems},
  \href{https://doi.org/10.1088/1126-6708/2007/09/120}{JHEP
    \textbf{09} (2007) 120}
       [\href{https://arxiv.org/abs/0708.4025}{0708.4025}].
\bibitem{Hubeny:2007xt}
  V.~E.~Hubeny, M.~Rangamani and T.~Takayanagi, {\it A Covariant
    holographic entanglement entropy proposal},
  \href{https://doi.org/10.1088/1126-6708/2007/07/062}{JHEP
    \textbf{07} (2007) 062}
  [\href{https://arxiv.org/abs/0705.0016}{0705.0016}].
\bibitem{Faulkner:2013ana}
  T.~Faulkner, A.~Lewkowycz and J.~Maldacena, {\it Quantum corrections
    to holographic entanglement entropy},
  \href{https://doi.org/10.1007/JHEP11(2013)074}{JHEP \textbf{11}
    (2013) 074} [\href{https://arxiv.org/abs/1307.2892}{1307.2892}].
\bibitem{Engelhardt:2014gca}
  N.~Engelhardt and A.~C.~Wall, {\it Quantum Extremal Surfaces:
    Holographic Entanglement Entropy beyond the Classical Regime},
  \href{https://doi.org/10.1007/JHEP01(2015)073}{JHEP \textbf{01}
    (2015) 073} [\href{https://arxiv.org/abs/1408.3203}{1408.3203}].
\bibitem{Penington:2019kki}
  G.~Penington, S.~H.~Shenker, D.~Stanford and Z.~Yang, {\it Replica
    wormholes and the black hole interior},
  \href{https://arxiv.org/abs/1911.11977}{1911.11977}.
\bibitem{Almheiri:2019qdq}
  A.~Almheiri, T.~Hartman, J.~Maldacena, E.~Shaghoulian and
  A.~Tajdini, {\it Replica Wormholes and the Entropy of Hawking
    Radiation}, \href{https://doi.org/10.1007/JHEP05(2020)013}{JHEP
    \textbf{05} (2020) 013}
  [\href{https://arxiv.org/abs/1911.12333}{1911.12333}].
\bibitem{Almheiri:2019hni}
  A.~Almheiri, R.~Mahajan, J.~Maldacena and Y.~Zhao, {\it The Page
    curve of Hawking radiation from semiclassical geometry},
  \href{https://doi.org/10.1007/JHEP03(2020)149}{JHEP \textbf{03}
    (2020) 149} [\href{https://arxiv.org/abs/1908.10996}{1908.10996}].
  \bibitem{Gautason:2020tmk}
  F.~F.~Gautason, L.~Schneiderbauer, W.~Sybesma and L.~Thorlacius,
  {\it Page Curve for an Evaporating Black Hole},
  \href{https://doi.org/10.1007/JHEP05(2020)091}{JHEP \textbf{05}
    (2020) 091}
   [\href{https://arxiv.org/abs/2004.00598}{2004.00598}].
\bibitem{Hartman:2020swn}
  T.~Hartman, E.~Shaghoulian and A.~Strominger, {\it Islands in
    Asymptotically Flat 2D Gravity},
  \href{https://arxiv.org/abs/2004.13857}{2004.13857}.
\bibitem{Maldacena:2001kr}
  J.~M.~Maldacena, {\it Eternal black holes in anti-de Sitter},
  \href{https://doi.org/10.1088/1126-6708/2003/04/021}{JHEP
    \textbf{04} (2003) 021}
  [\href{https://arxiv.org/abs/hep-th/0106112}{hep-th/0106112}].
\bibitem{Almheiri:2012rt}
  A.~Almheiri, D.~Marolf, J.~Polchinski and J.~Sully, {\it Black
    Holes: Complementarity or Firewalls?},
  \href{https://doi.org/10.1007/JHEP02(2013)062}{JHEP {\bf 1302}
    (2013) 062}
  [\href{https://arxiv.org/abs/1207.3123}{1207.3123}].
\bibitem{tHooft:1996rdg}
  G.~'t Hooft, {\it The Scattering matrix approach for the quantum
    black hole: An Overview},
  \href{https://doi.org/10.1142/S0217751X96002145}{Int.\ J.\ Mod.\ Phys.\ A
    {\bf 11} (1996) 4623}
  [\href{https://arxiv.org/abs/gr-qc/9607022}{gr-qc/9607022}].
\bibitem{Giddings:2009gj}
  S.~B.~Giddings and R.~A.~Porto, {\it The Gravitational S-matrix},
  \href{https://doi.org/10.1103/PhysRevD.81.025002}{Phys. Rev. D
    \textbf{81} (2010) 025002}
  [\href{https://arxiv.org/abs/0908.0004}{0908.0004}].
\bibitem{Bezrukov:2015ufa}
  F.~Bezrukov, D.~Levkov and S.~Sibiryakov, {\it Semiclassical
    S-matrix for black holes},
  \href{https://doi.org/10.1007/JHEP12(2015)002}{JHEP {\bf 1512}
    (2015) 002}
  [\href{https://arxiv.org/abs/1503.07181}{1503.07181}].
\bibitem{Berezin:1999nn}
  V.~Berezin, A.~Boyarsky and A.~Neronov, {\it On the Mechanism of
    Hawking radiation},
  Grav. Cosmol. \textbf{5} (1999) 16 
  [\href{https://arxiv.org/abs/gr-qc/0605099}{gr-qc/0605099}].
\bibitem{Parikh:1999mf}
  M.~K.~Parikh and F.~Wilczek, {\it Hawking radiation as tunneling},
  \href{https://doi.org/10.1103/PhysRevLett.85.5042}{Phys.\ Rev.\ Lett.\  {\bf
      85} (2000) 5042}
  [\href{https://arxiv.org/abs/hep-th/9907001}{hep-th/9907001}]. 
\bibitem{Bezrukov:2003tg}
  F.~Bezrukov and D.~Levkov, {\it Dynamical tunneling of bound systems
    through a potential barrier: complex way to the top},
  \href{https://doi.org/10.1134/1.1757681}{J. Exp. Theor. Phys. \textbf{98}
    (2004) 820}
  [\href{https://arxiv.org/abs/quant-ph/0312144}{quant-ph/0312144}].
\bibitem{Levkov:2007yn}
  D.~Levkov, A.~Panin and S.~Sibiryakov, {\it Unstable Semiclassical
    Trajectories in Tunneling},
  \href{https://doi.org/10.1103/PhysRevLett.99.170407}{Phys. Rev. Lett. \textbf{99}
    (2007) 170407}
  [\href{https://arxiv.org/abs/0707.0433}{0707.0433}].
\bibitem{Callan:1992rs}
  C.~G.~Callan, Jr., S.~B.~Giddings, J.~A.~Harvey and A.~Strominger,
  {\it Evanescent black holes},
  \href{https://doi.org/10.1103/PhysRevD.45.R1005}{Phys.\ Rev.\ D {\bf
      45} (1992) R1005}
  [\href{https://arxiv.org/abs/hep-th/9111056}{hep-th/9111056}].
\bibitem{Strominger:1994tn}
  A.~Strominger, {\it Les Houches lectures on black holes},
  \href{https://arxiv.org/abs/hep-th/9501071}{hep-th/9501071}.
\bibitem{Russo:1992ax}
  J.~G.~Russo, L.~Susskind and L.~Thorlacius, {\it The Endpoint of
    Hawking radiation},
  \href{https://doi.org/10.1103/PhysRevD.46.3444}{Phys. Rev. D
    \textbf{46} (1992) 3444}
  [\href{https://arxiv.org/abs/hep-th/9206070}{hep-th/9206070}].
\bibitem{Chung:1993rf}
  T.~D.~Chung and H.~L.~Verlinde, {\it Dynamical moving mirrors and
    black holes}, 
  \href{https://doi.org/10.1016/0550-3213(94)90249-6}{Nucl. Phys. B
    \textbf{418} (1994) 305}
  [\href{https://arxiv.org/abs/hep-th/9311007}{hep-th/9311007}].\
\bibitem{Strominger:1994xi}
  A.~Strominger and L.~Thorlacius, {\it Conformally invariant boundary
    conditions for dilaton gravity},
  \href{https://doi.org/10.1103/PhysRevD.50.5177}{Phys. Rev. D
    \textbf{50} (1994) 5177}
  [\href{https://arxiv.org/abs/hep-th/9405084}{hep-th/9405084}].
\bibitem{Das:1994yc}
  S.~R.~Das and S.~Mukherji, {\it Boundary dynamics in dilaton gravity},
  \href{https://doi.org/10.1142/S0217732394002938}{Mod. Phys. Lett. A
    \textbf{9} (1994) 3105}
  [\href{https://arxiv.org/abs/hep-th/9407015}{hep-th/9407015}].
\bibitem{Fitkevich:2017izc}
  M.~Fitkevich, D.~Levkov and Y.~Zenkevich, {\it Exact solutions and
    critical chaos in dilaton gravity with a boundary},
  \href{https://doi.org/10.1007/JHEP04(2017)108}{JHEP {\bf 1704}
    (2017) 108}
  [\href{https://arxiv.org/abs/1702.02576}{1702.02576}].
\bibitem{Teitelboim:1983ux}
  C.~Teitelboim, {\it Gravitation and Hamiltonian Structure in Two
    Space-Time Dimensions},
  \href{https://doi.org/10.1016/0370-2693(83)90012-6}{Phys. Lett. B
    \textbf{126} (1983) 41}.
\bibitem{Jackiw:1984je}
  R.~Jackiw, {\it Lower Dimensional Gravity,}
  \href{https://doi.org/10.1016/0550-3213(85)90448-1}{Nucl. Phys. B
    \textbf{252} (1985) 343}.
\bibitem{Cangemi:1992bj}
  D.~Cangemi and R.~Jackiw, {\it Gauge invariant formulations of
    lineal gravity},
  \href{https://doi.org/10.1103/PhysRevLett.69.233}{Phys. Rev. Lett. \textbf{69}
    (1992) 233}
  [\href{https://arxiv.org/abs/hep-th/9203056}{hep-th/9203056}].
\bibitem{Fitkevich:2020okl}
  M.~Fitkevich, D.~Levkov and Y.~Zenkevich, {\it Dilaton gravity with
    a boundary: from unitarity to black hole evaporation},
  \href{https://arxiv.org/abs/2004.13745}{2004.13745}.
\bibitem{Fiola:1994ir}
  T.~M.~Fiola, J.~Preskill, A.~Strominger and S.~P.~Trivedi,
  {\it Black hole thermodynamics and information loss in two-dimensions},
  \href{https://doi.org/10.1103/PhysRevD.50.3987}{Phys. Rev. D
    \textbf{50} (1994) 3987}
   [\href{https://arxiv.org/abs/hep-th/9403137}{hep-th/9403137}].
\bibitem{Myers:1994sg}
R.~C.~Myers,
{\it Black hole entropy in two-dimensions},
\href{https://doi.org/10.1103/PhysRevD.50.6412}{Phys. Rev. D
  \textbf{50} (1994) 6412}
[\href{https://arxiv.org/abs/hep-th/9405162}{hep-th/9405162}].
\bibitem{Hayward:1994dw}
  J.~D.~Hayward, {\it Entropy in the RST model},
  \href{https://doi.org/10.1103/PhysRevD.52.2239}{Phys. Rev. D
    \textbf{52} (1995) 2239}
  [\href{https://arxiv.org/abs/gr-qc/9412065}{gr-qc/9412065}].

\bibitem{Solodukhin:1995te}
  S.~N.~Solodukhin, {\it Two-dimensional quantum corrected eternal black hole,}
  \href{https://doi.org/10.1103/PhysRevD.53.824}{Phys.\ Rev.\ D {\bf
      53} (1996) 824}
  [\href{https://arxiv.org/abs/hep-th/9506206}{hep-th/9506206}].
\bibitem{Gibbons:1976ue}
  G.~W.~Gibbons and S.~W.~Hawking, {\it Action Integrals and Partition
    Functions in Quantum Gravity},
  \href{https://journals.aps.org/prd/abstract/10.1103/PhysRevD.15.2752}{Phys.\ Rev.\ D
    {\bf 15} (1977) 2752}.
\bibitem{Strominger:1996sh}
  A.~Strominger and C.~Vafa, {\it Microscopic origin of the
    Bekenstein-Hawking entropy},
  \href{https://doi.org/10.1016/0370-2693(96)00345-0}{Phys. Lett. B
    \textbf{379} (1996) 99}
  [\href{https://arxiv.org/abs/hep-th/9601029}{hep-th/9601029}].
\bibitem{deAlwis:1992emy}
  S.~de Alwis, {\it Quantization of a theory of 2-d dilaton gravity},
  \href{https://doi.org/10.1016/0370-2693(92)91219-Y}{Phys. Lett. B
    \textbf{289} (1992) 278}
  [\href{https://arxiv.org/abs/hep-th/9205069}{hep-th/9205069}].
\bibitem{LouisMartinez:1993cc}
  D.~Louis-Martinez and G.~Kunstatter, {\it Birkhoff's theorem in
    two-dimensional dilaton gravity},
  \href{https://doi.org/10.1103/PhysRevD.49.5227}{Phys.\ Rev.\ D {\bf
      49} (1994) 5227}.
\bibitem{Berezin:1987bc}
  V.~Berezin, V.~Kuzmin and I.~Tkachev, {\it Dynamics of Bubbles in
    General Relativity},
  \href{https://journals.aps.org/prd/abstract/10.1103/PhysRevD.36.2919}{Phys. Rev. D
    \textbf{36} (1987) 2919}.
\bibitem{Levkov:2007ce}
  D.~Levkov, A.~Panin and S.~Sibiryakov,
  {\it Complex trajectories in chaotic dynamical tunneling},
  \href{https://doi.org/10.1103/PhysRevE.76.046209}{Phys. Rev. E
    \textbf{76} (2007) 046209 }
   [\href{https://arxiv.org/abs/nlin/0701063}{nlin/0701063}].
\bibitem{Levkov:2008csa}
  D.~Levkov, A.~Panin and S.~Sibiryakov,
  {\it Signatures of unstable semiclassical trajectories in tunneling},
  \href{https://doi.org/10.1088/1751-8113/42/20/205102}{J. Phys. A
    \textbf{42} (2009) 205102}
  [\href{https://arxiv.org/abs/0811.3391}{0811.3391}].
\bibitem{Parikh:2004ih}
  M.~K.~Parikh,
  {\it A Secret tunnel through the horizon},
  \href{https://doi.org/10.1142/S0218271804006498}{Gen. Rel. Grav. \textbf{36}
    (2004) 2419} 
  [\href{https://arxiv.org/abs/hep-th/0405160}{hep-th/0405160}].
\bibitem{Banks:1992ba}
  T.~Banks, A.~Dabholkar, M.~R.~Douglas and M.~O'Loughlin, {\it Are
    horned particles the climax of Hawking evaporation?},
  \href{https://doi.org/10.1103/PhysRevD.45.3607}{Phys. Rev. D
    \textbf{45} (1992) 3607}
  [\href{https://arxiv.org/abs/hep-th/9201061}{hep-th/9201061}].
\bibitem{Russo:1992ht}
  J.~G.~Russo, L.~Susskind and L.~Thorlacius, {\it Black hole
    evaporation in (1+1)-dimensions},
  \href{https://doi.org/10.1016/0370-2693(92)90601-Y}{Phys. Lett. B
    \textbf{292} (1992) 13}
  [\href{https://arxiv.org/abs/hep-th/9201074}{hep-th/9201074}].
\bibitem{Thorlacius:1994ip}
  L.~Thorlacius, {\it Black hole evolution},
  \href{https://doi.org/10.1016/0920-5632(95)00435-C}{Nucl. Phys. B
    Proc. Suppl. \textbf{41} (1995) 245} 
  [\href{https://arxiv.org/abs/hep-th/9411020}{hep-th/9411020}].
\bibitem{Landauer:1994zz}
R.~Landauer and T.~Martin,
{\it Barrier interaction time in tunneling,}
\href{https://doi.org/10.1103/RevModPhys.66.217}{Rev. Mod. Phys. \textbf{66}
  (1994), 217}.


\end{thebibliography}
\end{document}